\documentclass[twoside,twocolumn,9pt]{article}
\usepackage{extsizes}
\usepackage[super,sort&compress,comma]{natbib} 
\usepackage[left=1.5cm, right=1.5cm, top=1.785cm, bottom=2.0cm]{geometry}
\usepackage{balance}
\usepackage[dvipsnames]{xcolor}
\usepackage{mathptmx}
\usepackage{sectsty}
\usepackage{graphicx} 
\usepackage{subfigure}
\usepackage{pifont}
\usepackage{lastpage}
\usepackage[format=plain,justification=justified,singlelinecheck=false,font={stretch=1.125,small,sf},labelfont=bf,labelsep=space]{caption}
\usepackage{float}
\usepackage{fancyhdr}
\usepackage{fnpos}
\usepackage[english]{babel}
\addto{\captionsenglish}{%
  
}
\usepackage{array}
\usepackage{droidsans}
\usepackage{charter}
\usepackage[T1]{fontenc}
\usepackage{setspace}
\usepackage[compact]{titlesec}
\usepackage[hide links]{hyperref}

\usepackage{natbib}
\usepackage{amsmath,amssymb,mathrsfs,mathtools}
\usepackage{soul}
\usepackage{bbding}
\usepackage[normalem]{ulem}
\usepackage[version=4]{mhchem}

\usepackage{siunitx}

\definecolor{cream}{RGB}{222,217,201}

\newcommand\nc{\newcommand}

\nc\cB{{\mathcal{B}}}
\nc\cC{{\mathcal{C}}}
\nc\cN{{\mathcal{N}}}
\nc\tcN{\tilde{{\cal{N}}}}
\nc\cG{{\mathcal{G}}}
\nc\cD{{\mathcal{D}}}
\nc\cU{{\mathcal{U}}}
\nc\cL{{\mathcal{L}}}
\nc\cT{{\mathcal{T}}}
\nc\cF{{\mathcal{F}}}
\nc\cK{{\mathcal{K}}}

\nc\ep{{\mathcal{\epsilon}}}
\nc{\pad}[2]{\frac{\partial #1}{\partial #2}}
\nc{\padn}[3]{\frac{\partial^{#3} #1}{\partial #2^{#3}}}

\usepackage[acronym]{glossaries}

\newacronym{NLC}{NLC}{nematic liquid crystal}
\newacronym{LSA}{LSA}{linear stability analysis}
\newacronym{ODE}{ODE}{ordinary differential equation}
\newacronym{PDE}{PDE}{partial differential equation}

\newacronym{nCB}{nCB}{cyanobiphenyl}
\newacronym{5CB}{5CB}{4-Cyano-4'-pentylbiphenyl}
\newacronym{6CB}{6CB}{4-cyano-4'-hexylbiphenyl}
\newacronym{7CB}{7CB}{4-Cyano-4'-heptylbiphenyl}
\newacronym{8CB}{8CB}{4-Octyl-4â-cyanobiphenyl}
\newacronym{5AB}{5AB}{tris(trimethylsiloxy)silane-ethoxycyanobiphenyl}

\usepackage[inline]{trackchanges}
\addeditor{ML}
\addeditor{LK}
\addeditor{LC}

\newsavebox{\astrutbox}
\sbox{\astrutbox}{\rule[-5pt]{0pt}{20pt}}

\nc{\vect}[1]{\mbox{\boldmath $#1$}}

\begin{document}

\pagestyle{fancy}
\thispagestyle{plain}
\fancypagestyle{plain}{
\renewcommand{\headrulewidth}{0pt}
}

\makeFNbottom
\makeatletter
\renewcommand\LARGE{\@setfontsize\LARGE{15pt}{17}}
\renewcommand\Large{\@setfontsize\Large{12pt}{14}}
\renewcommand\large{\@setfontsize\large{10pt}{12}}
\renewcommand\footnotesize{\@setfontsize\footnotesize{7pt}{10}}
\makeatother

\renewcommand{\thefootnote}{\fnsymbol{footnote}}
\renewcommand\footnoterule{\vspace*{1pt}%
\color{cream}\hrule width 3.5in height 0.4pt \color{black}\vspace*{5pt}} 
\setcounter{secnumdepth}{5}

\makeatletter 
\renewcommand\@biblabel[1]{#1}            
\renewcommand\@makefntext[1]%
{\noindent\makebox[0pt][r]{\@thefnmark\,}#1}
\makeatother 
\renewcommand{\figurename}{\small{Fig.}~}
\sectionfont{\sffamily\Large}
\subsectionfont{\normalsize}
\subsubsectionfont{\bf}
\setstretch{1.125} 
\setlength{\skip\footins}{0.8cm}
\setlength{\footnotesep}{0.25cm}
\setlength{\jot}{10pt}
\titlespacing*{\section}{0pt}{4pt}{4pt}
\titlespacing*{\subsection}{0pt}{15pt}{1pt}

\fancyfoot{}
\fancyfoot[LO,RE]{\vspace{-7.1pt}\includegraphics[height=9pt]{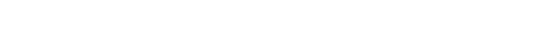}}
\fancyfoot[CO]{\vspace{-7.1pt}\hspace{13.2cm}\includegraphics{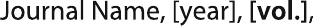}}
\fancyfoot[CE]{\vspace{-7.2pt}\hspace{-14.2cm}\includegraphics{head_foot/RF}}
\fancyfoot[RO]{\footnotesize{\sffamily{1--\pageref{LastPage} ~\textbar  \hspace{2pt}\thepage}}}
\fancyfoot[LE]{\footnotesize{\sffamily{\thepage~\textbar\hspace{3.45cm} 1--\pageref{LastPage}}}}
\fancyhead{}
\renewcommand{\headrulewidth}{0pt} 
\renewcommand{\footrulewidth}{0pt}
\setlength{\arrayrulewidth}{1pt}
\setlength{\columnsep}{6.5mm}
\setlength\bibsep{1pt}

\makeatletter 
\newlength{\figrulesep} 
\setlength{\figrulesep}{0.5\textfloatsep} 

\newcommand{\topfigrule}{\vspace*{-1pt}%
\noindent{\color{cream}\rule[-\figrulesep]{\columnwidth}{1.5pt}} }

\newcommand{\botfigrule}{\vspace*{-2pt}%
\noindent{\color{cream}\rule[\figrulesep]{\columnwidth}{1.5pt}} }

\newcommand{\dblfigrule}{\vspace*{-1pt}%
\noindent{\color{cream}\rule[-\figrulesep]{\textwidth}{1.5pt}} }

\makeatother

\twocolumn[
  \begin{@twocolumnfalse}
{\includegraphics[height=30pt]{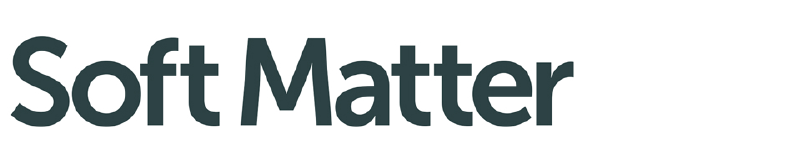}\hfill\raisebox{0pt}[0pt][0pt]{\includegraphics[height=55pt]{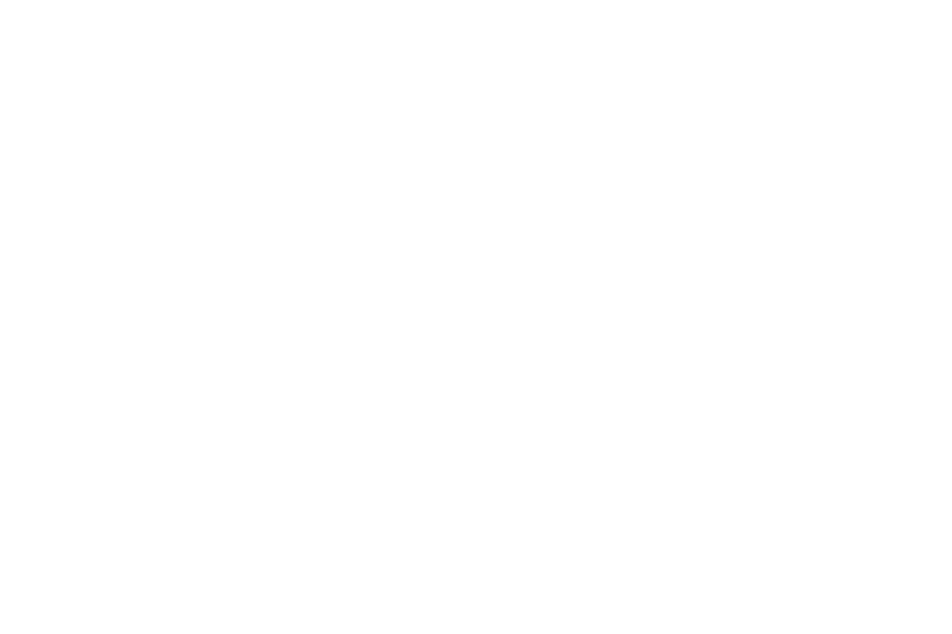}}\\[1ex]
\includegraphics[width=18.5cm]{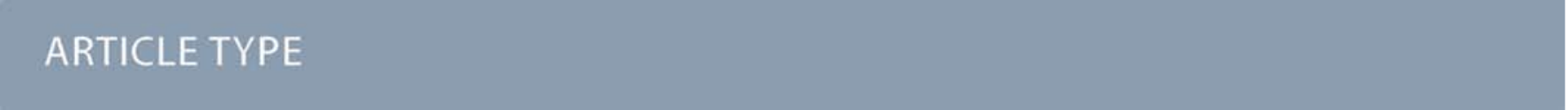}}\par
\vspace{1em}
\sffamily
\begin{tabular}{m{4.5cm} p{13.5cm} }

\includegraphics{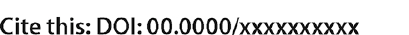} & \noindent\LARGE{\textbf{Effects of spatially-varying substrate anchoring on instabilities and dewetting of thin Nematic Liquid Crystal films
}} \\
\vspace{0.3cm} & \vspace{0.3cm} \\

 & \noindent\large{Michael-Angelo Y.-H. Lam,\textit{$^{a}$} Lou Kondic,\textit{$^{b}$} and Linda J. Cummings\textit{$^{b}$} } \\

\includegraphics{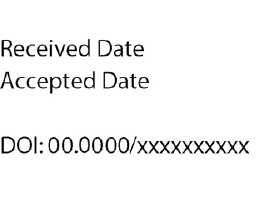} & \noindent\normalsize{
Partially wetting nematic liquid crystal (NLC) films on substrates are unstable to dewetting-type instabilities due to destablizing solid/NLC interaction forces.  These instabilities are modified by the nematic nature of the films, which influences the effective solid/NLC interaction.  In this work, we focus on the influence of imposed substrate anchoring on the instability development. The analysis is carried out within a long-wave formulation 
based on the Leslie-Ericksen description of NLC films.  Linear stability analysis of the resulting equations shows that some features of the instability, such as emerging wavelengths, may not be influenced by the imposed substrate anchoring. Going further into the nonlinear regime, considered via large-scale GPU-based simulations, shows however that nonlinear effects may play an important role, in particular in the case of strong substrate anchoring anisotropy. Our simulations show that instability of the film develops in two stages: the first stage involves formation of ridges that are perpendicular to the local anchoring direction; and the second involves breakup of these ridges and formation of drops, whose final distribution is influenced by the anisotropy imposed by the substrate.  Finally, we 
show that imposing more complex substrate anisotropy patterns allows us to reach basic understanding of the influence of substrate-imposed defects in director 
orientation on the instability evolution.  
} \\

\end{tabular}

 \end{@twocolumnfalse} \vspace{0.6cm}     ]


\renewcommand*\rmdefault{bch}\normalfont\upshape
\rmfamily
\section*{}
\vspace{-1cm}


\footnotetext{\textit{$^{a}$~U. S. Army Engineer Research and Development Center, Coastal and Hydraulics Laboratory, 3909 Halls Ferry Road, Vicksburg, MS 39180, USA. E-mail: michaelangelo.yh.lam@gmail.com }}
\footnotetext{\textit{$^{b}$~Department of Mathematical Sciences and Center for Applied Mathematics and Statistics, New Jersey Institute of Technology, Newark, New Jersey 07102, USA. E-mail: kondic@njit.edu, linda.cummings@njit.edu}}






\section{Introduction}


\Gls{NLC} is one of several possible liquid crystalline states of matter, intermediate between a solid (crystal) and a liquid, that can exist.  Typically, the molecules of \gls{NLC}s are rod-like and interact electrostatically, which leads to them exhibiting short-range directional ordering, an elastic response under deformation, and anisotropic viscosity when they flow. At interfacial boundaries, whether free or rigid, NLC molecules typically have a preferred orientation, a phenomenon known as {\it anchoring}. Due to the electrostatic interactions between molecules, anchoring affects strongly molecular orientation within the bulk of the NLC, and how the sample flows and deforms.  
While NLC films are interesting in their own right, we note that such films, as well as the mathematical models used for their description, share many common aspects with active fluids, which often involve rod-like particles that may attain nematic order.  We refer the reader to excellent reviews discussing a number of related active-matter systems~\cite{marchetti_rmp_2013,saintillan_arfm_2018}, as well as to 
specific recent research papers that focus on the relation between active and passive anisotropic 
films~\cite{ramaswamy_prl_2009,joanny_jfm_2012,thiele_pre_2020,blow_sm_2017}.

A fairly extensive experimental literature exists on the behavior of thin NLC films with a free surface (see, e.g. works by Cazabat {\it et al.}\cite{Cazabat2011}, Delabre {\it et al.}\cite{Delabre2009}, Herminghaus {\it et al.}\cite{Herminghaus1998}, and van Effenterre \& Valignat~\cite{Effenterre2003} among many others). In all of these works it is thought that the anchoring is spatially homogeneous; typically homeotropic (molecules perpendicular to interface) at the free surface, and degenerate planar at the substrate (molecules align parallel to the substrate in the orientation that minimizes the bulk elastic energy). Our earlier theoretical work~\cite{Lam2018,lam_jcp_2019} considered this situation in detail from both an analytical and numerical perspective, presenting a model that could replicate the instability and dewetting phenomena observed in the experiments. 
Non-degenerate (directional), spatially-varying substrate anchoring has been considered experimentally, but primarily within confined rigid geometries (a sandwich configuration), with a view to engineering multistable Liquid Crystal Display (LCD) devices~\cite{apl_yokoyama2001,jcp_schoen2012,nature_yokoyama2002}; 
some more recent works also consider the details of anchoring effects for active films~\cite{blow_sm_2017}. We are however unaware of experimental or theoretical work that studies the effects of nonuniform substrate anchoring on flow, spreading and instability of free surface NLC films.  Perhaps the most directly relevant work of which we are aware, which takes account of the effects of local molecular orientation on flow, is that of Forest {\it et al.}~\cite{forest_sm_2012}
, who use a diffuse-interface framework within the Doi-Hess kinetic theory for liquid crystal polymer droplets to study NLC droplets computationally under imposed shear in the presence of internal defects. 

In this paper we present a minimal model for the flow and dewetting of thin (nanoscale) films of NLC on a flat substrate at which the strong planar anchoring is allowed to vary spatially. Free surface anchoring is assumed to be weak and homeotropic, following our earlier work~\cite{Lam2018,Lin2013}. The model is based on the Leslie-Ericksen theory for NLCs, and accounts for van der Waals interactions between the NLC and the substrate, in addition to the bulk elasticity and surface energy contributions. In the spirit of formulating the simplest model capable of capturing the key physics, we neglect additional surface effects such as interfacial dissipation that could play a minor role in influencing the dynamics. This is also our motivation for choosing the Leslie-Ericksen model over a more comprehensive (but complicated) theory such as $Q$-tensor theory: the model we derive is much more tractable (analytically and numerically) than would be possible by starting from alternative models. We refer the reader to the review by Rey~\cite{rey_sm_2007} for an overview of works that use complementary approaches to modeling thin NLC films (such as the Landau--de Gennes formulation); see also more recent relevant work by Rey \& Herrera-Valencia on modeling the isotropic-to-nematic transition in a dynamic wetting context using this approach~\cite{rey_sm_2010}, as well as the above-referenced work by Forest {\it et al.}~\cite{forest_sm_2012}. 

In the present work, we focus particularly on the effect that local directionality of substrate anchoring has on the evolution of the overlying film, and the droplet patterns obtained at large times after film breakup. Unidirectional anchoring is considered first by way of illustration, being sufficiently simple that linear stability analysis can be carried out and used to predict results. Large-scale simulations are presented using an ADI scheme implemented on a GPU, first for the unidirectional anchoring case, and then for more complex anchoring patterns. Our results reveal that local directionality of substrate anchoring can affect significantly the patterns that emerge when a NLC film destabilizes and breaks up. It is our hope that future experimental work will be able to confirm our model predictions. 

The remainder of our manuscript is organized as follows.  In Sec.~\ref{sec:model}, we present the asymptotic
model that we will use for describing evolution of NLC films.  The presentation focuses in particular 
on the inclusion of spatially-dependent substrate anchoring, since other aspects of the model can be found 
in our earlier work~\cite{Lam2018,Lin2013}.   
Section~\ref{sec:LSA} discusses linear stability analysis of NLC films in a few simple setups that outline the influence that nonuniform substrate anchoring 
is expected to have on instability development.  The nonlinear stage of the evolution is considered
via fully-implicit large-time simulations that are presented in Sec.~\ref{sec:simulation}.  Here we show that nonlinear effects play an important role in the instability development and resulting 
pattern formation. Section~\ref{sec:conclusions} is devoted to a summary of key findings and discussion of possible future research directions. 

\section{Model Description} \label{sec:model}

\begin{figure}[t!]
	\centering
	\includegraphics[width=0.35\textwidth]{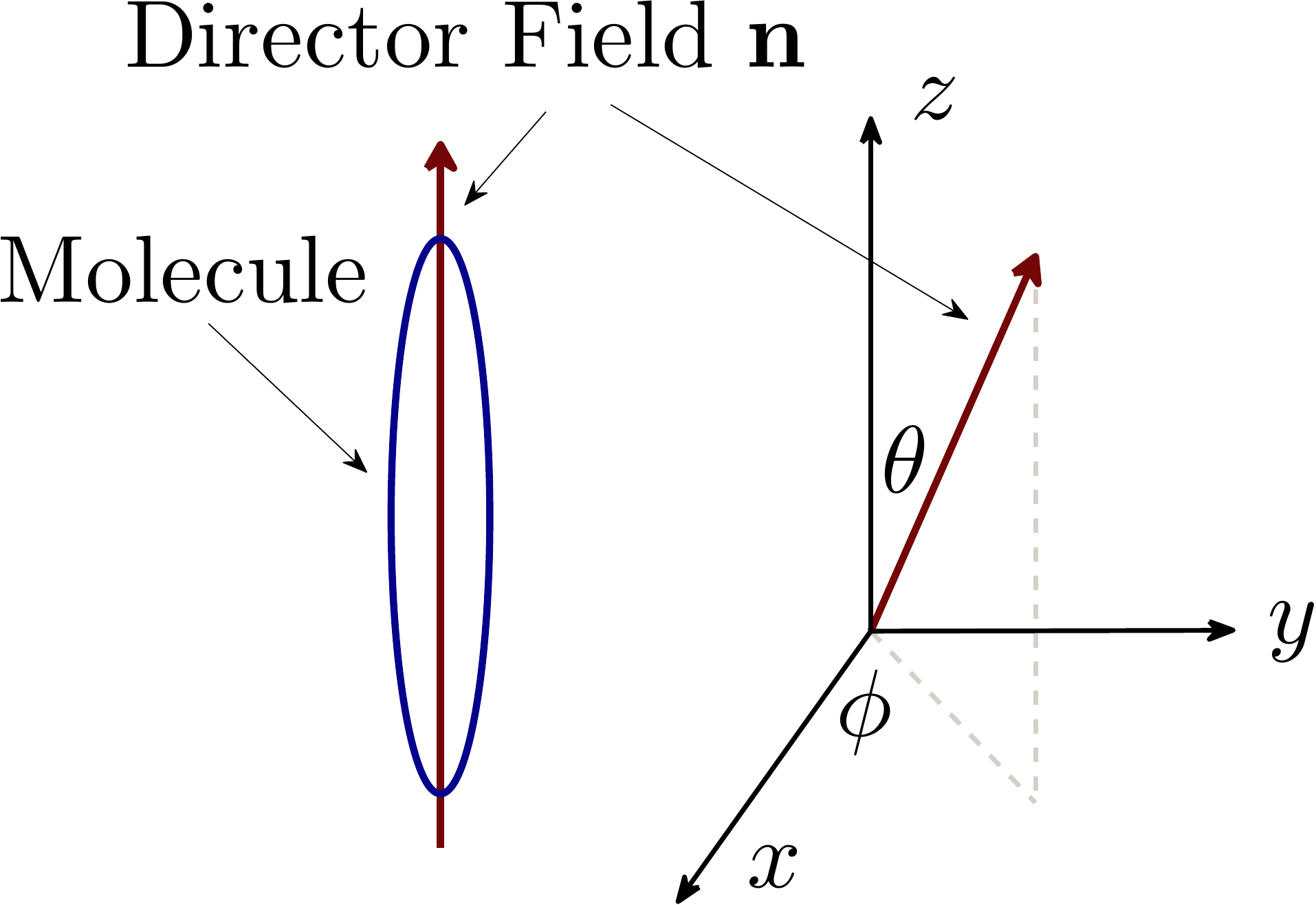}
	\caption{Schematic of director field (red arrow) relative to liquid crystal molecule (blue ellipse), and its description via spherical polar angles. } 
	\label{fig:DirectorField}
\end{figure}

 Associated with each rod-like NLC molecule is an electrical dipole moment, the interactions between which lead to an elastic response under deformation, resulting in short-range directional ordering of the molecules.  To describe the flow of \gls{NLC}, in addition to the velocity field, $\hat{\mathbf{v}}=(\hat{v}_1,\hat{v}_2,\hat{v}_3) = (\hat{v},\hat{u},\hat{w})$, one must also track the orientation of the \gls{NLC} molecules, modeled by a director field, $\mathbf{n}=(n_1,n_2,n_3)$, a unit vector representing the local average orientation. (Throughout this paper, hatted variables denote dimensional quantities and unhatted variables denote dimensionless ones.) The unit vector is typically aligned with the long axis of the \gls{NLC} molecules, see Figure~\ref{fig:DirectorField}, and it is often convenient to characterize the director field in terms of its polar angle, $\theta$, and azimuthal angle, $\phi$, considered as functions of Cartesian space variables $(\hat{x},\hat{y},\hat{z})=(\hat{x}_1,\hat{x}_2,\hat{x}_3)$, i.e, $\mathbf{n}=(\sin \theta\cos\phi,\sin \theta\sin \phi , \cos\theta)$.
  
The flow of \gls{NLC}s may be modeled using the Leslie-Ericksen (LE) equations~\cite{Leslie1979}, an extension of the Navier-Stokes equations, with an additional equation modeling conservation of energy. For brevity, we do not discuss the details of the derivation of the LE model, but note that it is based on four conservation laws: energy, linear momentum, angular momentum, and mass.  Assuming isothermal static deformations, the conservation of energy and angular momentum equations may be combined.   In addition, to model the anisotropic viscosity, the viscous stress tensor, $\hat{t}_{ij}$, is assumed to be a linear function of $\hat{e}_{ij}$, the symmetric rate of strain tensor; and $\hat{N}_j$, the rotation of the director field in the reference frame of moving antisymmetric deformations (characterized by the antisymmetric strain rate tensor, $\hat{\omega}_{ij}$). The quantity $\hat{N}_j$ may be interpreted as the additional rotational velocity component of the director field due to the (external) elastic response, which is separate from rotation imparted by the (internal) velocity field.   These quantities are defined as  
\begin{equation}
	\hat{e}_{ij} = \frac{1}{2}\left(\pad{\hat{v}_i}{\hat{x}_j} + \pad{\hat{v}_j}{\hat{x}_i} \right) \; , \;
	\hat{N}_i = \pad{n_i}{\hat{t}}-\hat{\omega}_{ij}n_j \; , \;
	\hat{\omega}_{ij} = \frac{1}{2}\left(\pad{\hat{v}_i}{\hat{x}_j} - \pad{\hat{v}_j}{\hat{x}_i} \right) \;.
\end{equation}

Under these broad assumptions, sixteen coefficients are required to define the viscous stress tensor; however, applying the laws of thermodynamics this number may be reduced to just six, $\hat{\alpha}_i$, $i=1,2,3,4,5,6$, simplifying the viscous stress tensor to
\begin{equation}
	\hat{t}_{ij} = \hat{\alpha}_1 n_k n_p \hat{e}_{kp} n_i n_j +
				\hat{\alpha}_2 \hat{N}_i n_j +
				\hat{\alpha}_3 \hat{N}_j n_i +
				\hat{\alpha}_4 \hat{e}_{ij} +
				\hat{\alpha}_5 \hat{e}_{ik} n_k n_j +
				\hat{\alpha}_6 \hat{e}_{jk} n_k n_i \;.
\end{equation}
Using the Onsager relation, $\hat{\alpha}_2 + \hat{\alpha}_3 = \hat{\alpha}_6 - \hat{\alpha}_5$, further reduces the number of independent coefficients to five.  Note that $\hat{\alpha}_4$ here plays the role of the viscosity coefficient for an isotropic Newtonian fluid. 

\begin{figure}[t!]
	\centering
	\subfigure[]{
		\includegraphics[height=0.18\textwidth]{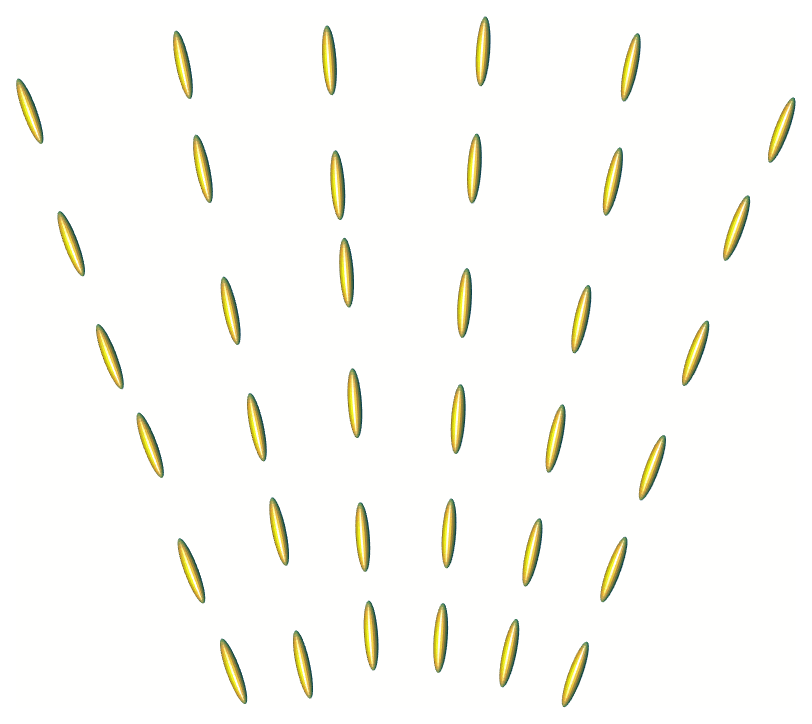} 
		\label{fig:Splay} 
	}
	\subfigure[]{
		\includegraphics[height=0.18\textwidth]{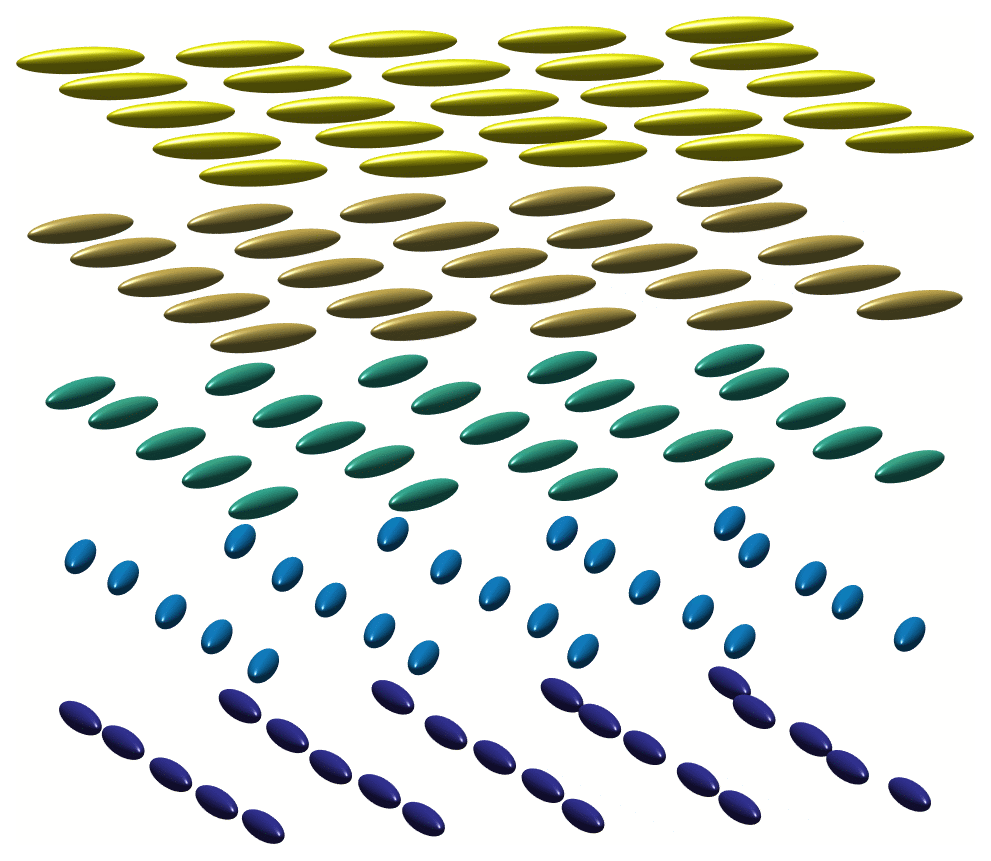}
		\label{fig:Twist} 
	}
	\subfigure[]{
		\includegraphics[height=0.18\textwidth]{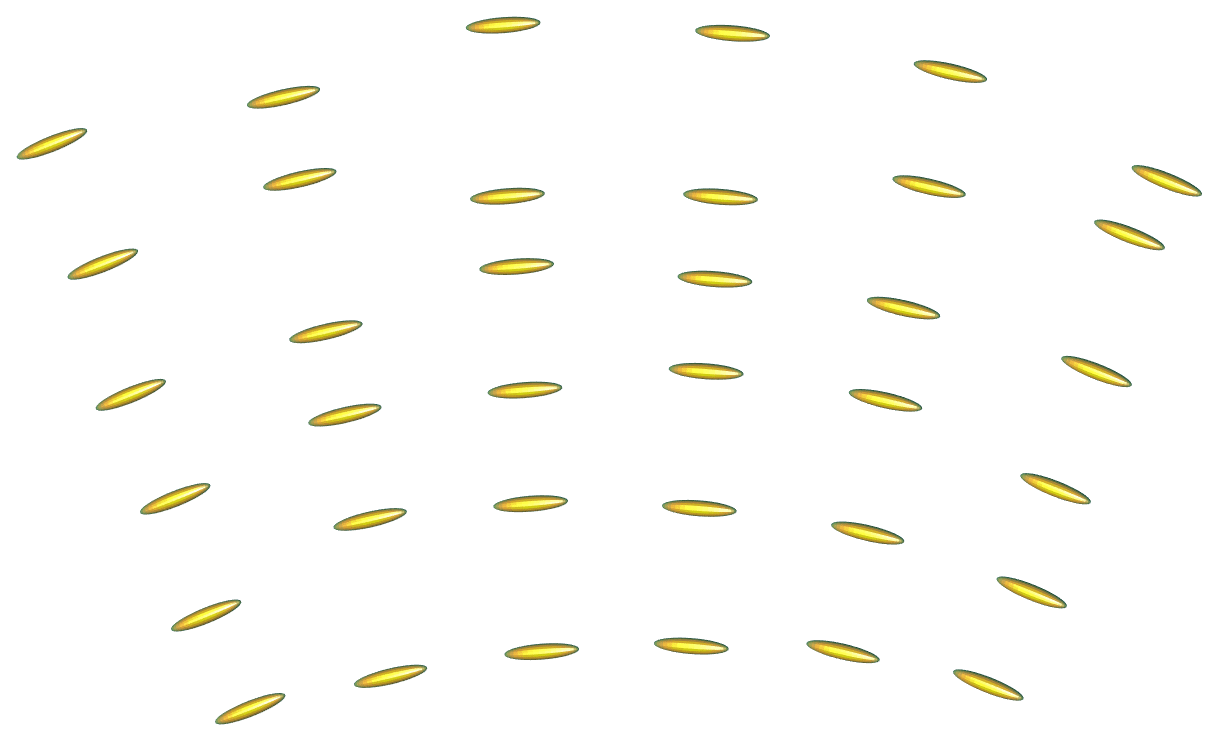} 
		\label{fig:Bend} 
	}
	\caption{Schematics of the deformation types modeled in the bulk elastic energy (\ref{eq:ElasticEnergy}) representing pure a) splay, b) twist, and c) bend.	}
	\label{fig:Deformations} 
\end{figure}

In addition to the internal forces captured by the stress tensor, the LE equations model external body forces on the director field,
\begin{equation}
	\hat{G}_i = \hat{\gamma}_1 \hat{N}_i + \hat{\gamma}_2 \hat{e}_{ij} n_j \; ,
\end{equation}
where $\hat{\gamma}_1=\hat{\alpha}_2 - \hat{\alpha}_3$ is the rotational viscosity (giving rise to a force on the NLC molecules due to rotational flow) and $\hat{\gamma}_2 = \hat{\alpha}_5 - \hat{\alpha}_6$ is the irrotational viscosity (giving a shear force on the molecules).  To model the elastic response of the \gls{NLC}, the bulk (Frank) elastic energy $\hat{W}$ is assumed to be a positive definite quadratic function of spatial derivatives of the director field. Specifically,
\begin{equation} \label{eq:ElasticEnergy}
	\hat{W} =\frac{1}{2} 
	\left[ 
		\hat{K}_1 \left( \hat{\nabla} \cdot \mathbf{n} \right)^2 +
		\hat{K}_2 \left(\mathbf{n} \cdot \hat{\nabla} \times \mathbf{n} \right)^2 +
		\hat{K}_3 \left| \mathbf{n} \times \hat{\nabla} \times \mathbf{n} \right|^2
	\right] \;,
\end{equation}
representing splay ($\hat{K}_1$), twist ($\hat{K}_2$) and bend ($\hat{K}_3$) deformation, see Figure~\ref{fig:Deformations}. It is common to use the so-called one-constant approximation~\cite{Delabre2009,Rey2008,Effenterre2003,Ziherl2003}, $\hat{K}=\hat{K}_1=\hat{K}_2=\hat{K}_3$, a practice we also follow. 

For an incompressible fluid, the LE equations are given by 
\begin{eqnarray}
	\pad{\hat{W}}{n_i} + \pad{}{\hat{x}_j} \left( \pad{\hat{W}}{\hat{n}_{i,j}} \right) - \hat{G}_i & = & 0, \quad \hat{n}_{i,j} = \pad{n_i}{\hat{x}_j} \;, \label{eq:LEConsEnergy} \\
	\pad{\hat{p}}{\hat{x}_i} + \pad{\hat{W}}{\hat{x}_i} + \hat{G}_j \pad{n_j}{\hat{x}_i} + \pad{\hat{\tau}_{ij}}{\hat{x}_j
	} & = & 0 \;, \label{eq:LEConsMomentum} \\
	\pad{\hat{v}_i}{\hat{x}_i} & = & 0 \label{eq:LEConsMass} \;,
\end{eqnarray}
respectively representing the combined conservation of energy and angular momentum (for isothermal static deformations of the NLC), conservation of linear momentum, and conservation of mass for an incompressible fluid.  For brevity, we will refer to equation (\ref{eq:LEConsEnergy}) as the energy equation.

\subsection{Nondimensionalization and scalings} \label{sec:Nondimensionalization}

To nondimensionalize the LE equations (\ref{eq:LEConsEnergy})--(\ref{eq:LEConsMass}) for the ``thin film'' scenarios we seek to describe, we define five scaling parameters: $\hat H$, a representative film thickness; $\hat L$, the lengthscale of variations in the plane of the substrate, $(\hat x,\hat y)$; $\hat T_{\rm f}$, the timescale for fluid flow;  $\hat T_{\rm r}$,  the timescale of elastic reorientation of NLC molecules; and $\hat \mu=\hat \alpha_4$, the representative viscosity corresponding to the isotropic Newtonian fluid case. In addition, we define the film aspect ratio, $\delta=\hat H/\hat L$, and assume $\delta\ll 1$ (the long wave approximation).  The values assigned to the scaling parameters are chosen based on the experiments of Herminghaus {\it et al.}~\cite{Herminghaus1998} and Cazabat {\it et al.}~\cite{Cazabat2011} for thin films of NLC, 
$\hat{H} = 100$ nm, $\hat{L} = 10$ $\mu$m, and $\hat{T}_{\rm f} = 1$ s.
We will see that viscosity appears in our final model via a single dimensionless parameter $\eta$, a ratio of a linear combination of other system viscosities to $\hat{\alpha}_4$. We will discuss its value later. As discussed below, provided $\hat{T}_{\rm r}\ll \hat{T}_{\rm f}$, the exact value of $\hat{T}_{\rm r}$ is irrelevant for our model.

We note that in dewetting experiments, a so-called ``forbidden range'' of film thicknesses (10 nm to 100 nm) is observed, within which NLC films are observed to be unstable, as well as a minimum film thickness, corresponding to a trilayer of molecules just a few nanometers thick~\cite{Cazabat2011}. We therefore define $\hat\beta$ to be the upper thickness threshold for film stability and $\hat b$ as the minimum film thickness (which we will refer to as the equilibrium film thickness). Consistent with available data for NLC systems, we set these values to 
$\hat{\beta} = 100$ nm and $\hat{b} = 1$ nm. 

\subsection{Energetics: Weak Anchoring Model}

Scaling quantities as follows,
\begin{equation} \label{eq:NLC_scales2}
	(\hat{x},\hat{y}, \hat{z})  =  \hat{L} (x,y, \delta z) \;, \quad
	(\hat{v},\hat{u}, \hat{w})  =  \frac{\hat{L}}{\hat{T}_{\rm f}} (u,v, \delta w) \;, \quad
	\hat{t} = \hat{T}_{\rm f} t \;, \quad
	\hat{\alpha}_i = \hat{\alpha}_4 \alpha_i \; ,
\end{equation}
and assuming further that the timescale of elastic reorientation is much faster than that of fluid flow, $\hat{T}_{\rm r} \ll \hat{T}_{\rm f}$, the (dimensionless) external body forces, $\hat{G}_i$, in the energy equation (\ref{eq:LEConsEnergy}), are seen to be negligible. To leading order then, equation (\ref{eq:LEConsEnergy}) decouples from equations (\ref{eq:LEConsMomentum}) and (\ref{eq:LEConsMass}), and reduces simply to an equation for the director field, $\mathbf{n}$: the problem of minimizing the free energy of the system~\cite{Cummings2004,Lam2018,Lin2013} with respect to variations in the polar angle $\theta$ and the azimuthal angle $\phi$. Upon solving, $\theta$ and $\phi$ are determined to be of the form
\begin{equation} \label{eq:GeneralDirectorField}
	\theta(x,y,z,t) = c_1(x,y,t)z + c_2(x,y,t) \quad \textrm{and} \quad 
	\phi(x,y,t) = c_3(x,y,t) \; ,
\end{equation}
where $c_1(x,y,t)$, $c_2(x,y,t)$, and $c_3(x,y,t)$ are independent of $z$ and must be chosen to satisfy appropriate ``anchoring'' boundary conditions at both the substrate and the free surface.  The director field thus bends but does not twist across the layer: the degree of bending is determined by the imposed anchoring conditions, discussed below, thus the director field is a function of the instantaneous fluid height, $z=h(x,y,t)$.  

\begin{figure}[t!]
	\centering
	\begin{minipage}{.45\textwidth}
		\centering
		\subfigure[]{
			\includegraphics[width=0.4\textwidth]{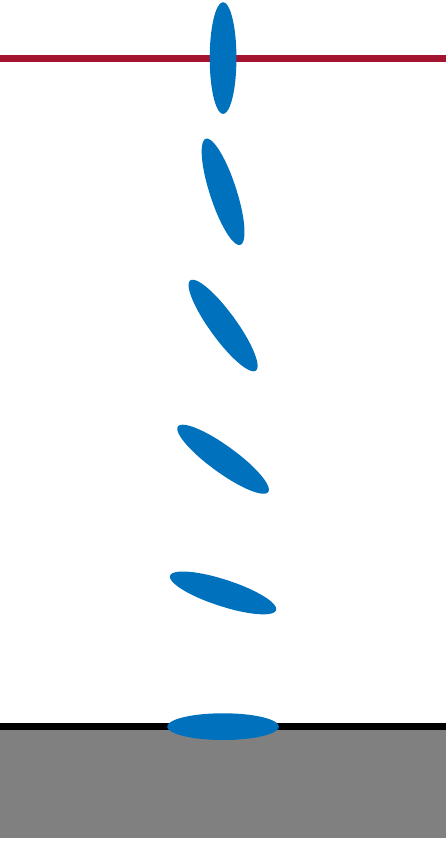}
			\label{fig:StrongAnchoringThick} 
		}
		\subfigure[]{
			\includegraphics[width=0.4\textwidth]{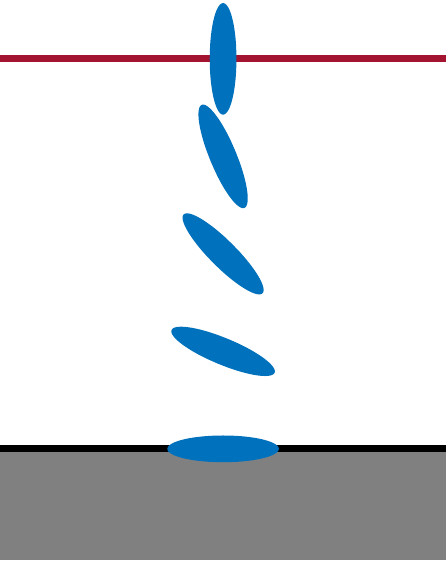} 
			\label{fig:StrongAnchoringThin} 
		}
		\captionof{figure}{Schematic of molecular orientation in NLC layer with strong homeotropic free surface anchoring and strong planar substrate anchoring for a) a thick film and b) a thin film. } 
		\label{fig:StrongAnchoring}
	\end{minipage}
	\quad
	\begin{minipage}{.45\textwidth}
		\centering
		\subfigure[]{
			\includegraphics[width=0.4\textwidth]{figures/AnchoringThick.pdf}
			\label{fig:WeakAnchoringThick} 
		}
		\subfigure[]{
			\includegraphics[width=0.4\textwidth]{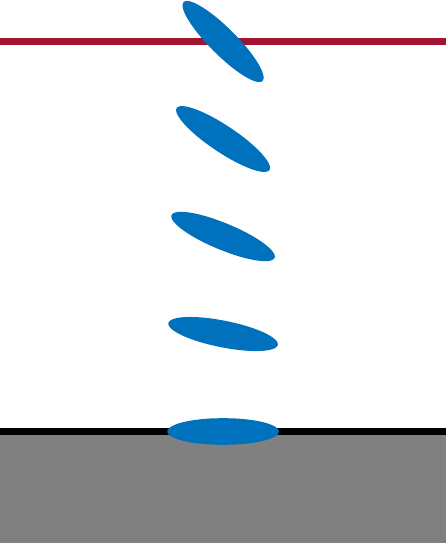} 
			\label{fig:WeakAnchoringThin} 
		}
		\captionof{figure}{ Schematic of molecular orientation in NLC layer with weak homeotropic free surface anchoring and strong planar substrate anchoring for a) a thick film and b) a thin film. } 
		\label{fig:WeakAnchoring}
	\end{minipage}
\end{figure}

At an interface, \gls{NLC} molecules typically have a preferred orientation, often called the anchoring condition. In many experiments the substrate is treated, either chemically or mechanically, such that molecules align in the plane of the substrate (planar anchoring), while at the free surface molecules often align perpendicular to the surface (homeotropic anchoring, a special case of conical anchoring in which the molecules prefer to orient on the surface of a cone of given angle with axis perpendicular to the free surface), see Figure~\ref{fig:StrongAnchoring}. This situation, where the director is required to adopt different orientations at opposite sides of a layer, is referred to as ``antagonistic anchoring''.   For relatively thick films the director can bend across the film to accommodate the two different anchoring conditions (see Figure~\ref{fig:StrongAnchoringThick}). However, for very thin films or close to a contact line, strict imposition of the antagonistic conditions can lead to large energy penalties in the bulk of the fluid due to the rapid spatial variations that result in the director field (see Figure~\ref{fig:StrongAnchoringThin}).   To alleviate this issue, we first note that in practice, anchoring strength at the substrate is usually stronger than at the free surface, therefore, we impose strong planar anchoring on the substrate and implement a weak free surface anchoring model as used in our previous work~\cite{Lam2014,Lam2015,Lam2018,lam_jcp_2019}, which allows the polar angle at the free surface to relax from the homeotropic state (valid for thick films) to the planar state, as the film thickness $h$ approaches the equilibrium thickness (see Figure~\ref{fig:WeakAnchoring}). 

Since the azimuthal director angle $\phi$ is found to be independent of the vertical coordinate $z$, we assume it is entirely determined by the (strong) substrate anchoring, which we allow to vary spatially: $\phi = \phi_{\rm S} (x,y)$.  Our particular focus in this paper is to investigate how such imposed spatially-varying substrate anchoring can influence the evolution of the overlying NLC layer. In practice, inhomogeneous anchoring could be achieved by a variety of techniques, including simple mechanical means such as rubbing a surface with a cloth in a prescribed direction. Therefore, we will prescribe $\phi_{\rm S} (x,y)$ as a boundary condition in our model.

Under these modeling assumptions the director angles are given by
\begin{equation} \label{eq:WeakAnchoringDirectorField}
	\theta(x,y,z) = \theta_{\rm S} + (\theta_{\rm F}-\theta_{\rm S}) \frac{m(h)}{h}z  \quad \textrm{and} \quad 
	\phi(x,y) = \phi_{\rm S}(x,y) \; ,
\end{equation}
where $h=h(x,y,t)$ is the free surface height, $m(h)$ is a function that captures the details of the weak anchoring, and for generality, we use the subscripts S and F to denote the prescribed anchoring angles at the substrate and free surface, respectively (for the specific case discussed above, $\theta_{\rm S}=\pi/2$ while $\theta_{\rm F}=0$). Following our earlier work~\cite{Lam2014,Lam2015,Lam2018,lam_jcp_2019} the weak anchoring function is chosen to be of the form
\begin{equation} \label{eq:mh_model}
m(h) = g(h) \frac{h^2}{h^2+\beta^2} \; ; \quad g(h)= \frac{1}{2}\left[ 1 + \tanh \left( \frac{h-2b}{w}\right) \right]\;,
\end{equation}
where $\beta =\hat{\beta}/\hat{H}$ is a film thickness at which bulk elastic energies are comparable to surface anchoring energies and $g(h)$ is a `cutoff' function (with width controlled by $w$, where we take $w=0.05$ throughout this work) that forces the free surface anchoring to match that of the substrate for film thicknesses close to the equilbrium film thickness $b=\hat{b}/\hat{H} \ll \beta$. For $h\gg\beta$ we have $m(h)\approx 1$, corresponding to the preferred free surface anchoring angle being attained ($\theta(x,y,h)\approx \theta_{\rm F}$ in (\ref{eq:WeakAnchoringDirectorField})). However, as $h\to b$ (the minimum thickness permitted by the governing partial differential equation, given in (\ref{eq:GovEqn}) below, on account of the disjoining pressure term specifed in (\ref{eq:DisjPressNLC})), the function $m(h)$ becomes very small and $\theta(x,y,h)\approx \theta_{\rm S}$ in (\ref{eq:WeakAnchoringDirectorField}).

\subsection{Long Wave Equation}
We now briefly discuss the long wave model that results from equations (\ref{eq:LEConsMomentum}) and (\ref{eq:LEConsMass}) with the scalings of Sec.~\ref{sec:Nondimensionalization}, referring the reader to Lam {\it et al.}~\cite{Lam2014} for full details. With the expressions given in equation (\ref{eq:WeakAnchoringDirectorField}) for the director angles $\theta$ and $\phi$, we substitute in equations (\ref{eq:LEConsMomentum}) to obtain partial differential equations that depend only on the free surface height, $h$; the velocity field, $\mathbf{v}$; and the pressure, $p$. 

\begin{figure}[t!]
	\centering
	\includegraphics[width=0.45\textwidth]{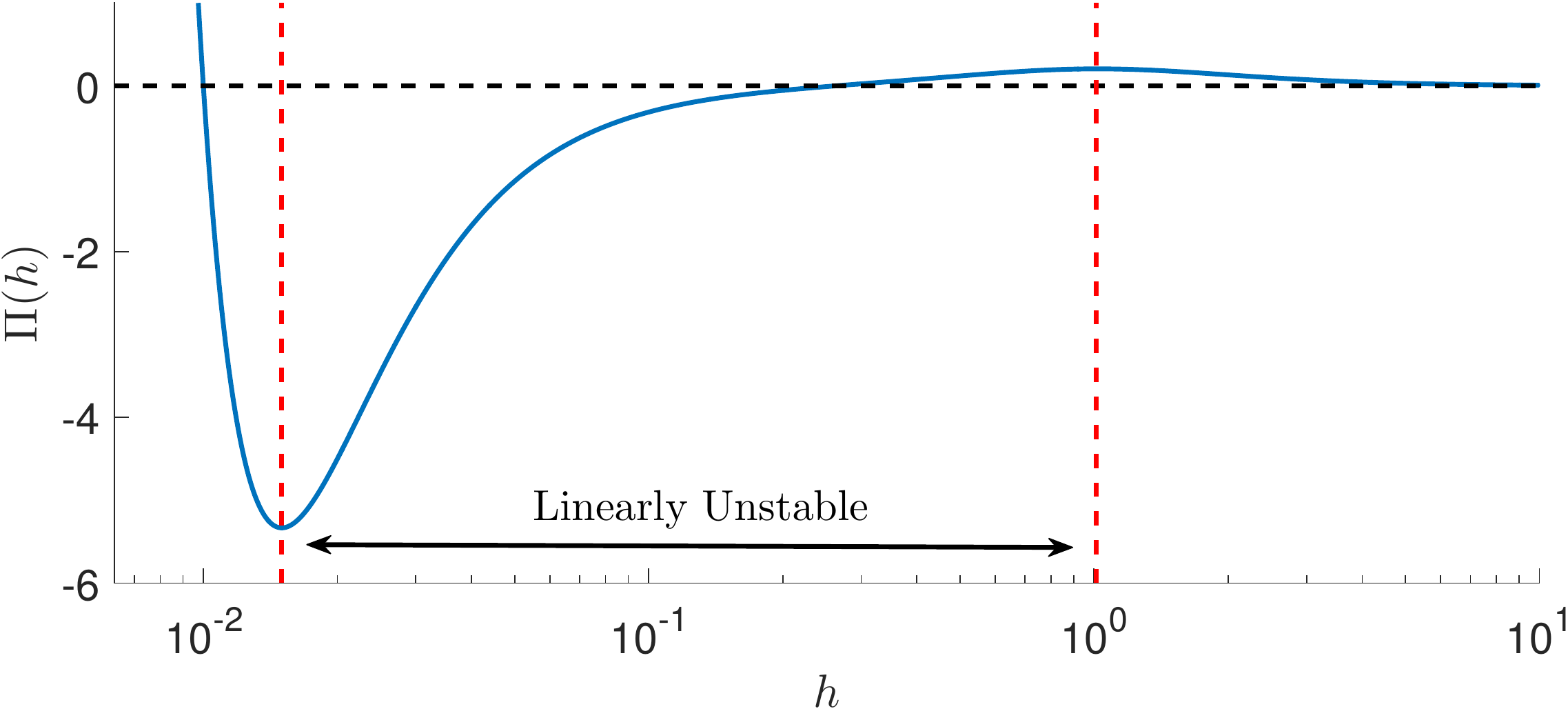} 
	\caption{Plot of the disjoining pressure eqn (\ref{eq:DisjPressNLC}) as a function of the film height, for the parameter values of (\ref{eq:NLC_paras}). Region between the vertical dashed red lines denotes the linearly unstable regime.	} 
	\label{fig:DisjoiningPressure}
\end{figure}

Under the long wave approximation, the leading-order transverse momentum equation ($z$-component of equation (\ref{eq:LEConsMomentum})) may be solved for the pressure on application of the normal stress balance boundary condition, while the leading-order in-plane momentum equations can be integrated over the film height from $z=0$ to $z=h(x,y,t)$, giving (after application of the usual no-slip and tangential stress boundary conditions) a fourth-order partial differential equation for the evolution of the free surface height. Motivated by previous work~\cite{Lam2018,lam_jcp_2019}, we choose to express the resulting long-wave equation in terms of the variational or gradient dynamics formulation~\cite{Mitlin1993,ThAP2016prf}, in which the evolution of the free surface height is given by

\begin{equation} \label{eq:GovEqn}
	\pad{h}{t} + \nabla \cdot \left[ \mathbf{Q}(h,\phi) \nabla \left( \frac{\delta E}{\delta h} \right) \right] = 0 \;,
\end{equation}
where $\mathbf{Q}$ is the mobility function and $E$ is total interfacial energy (Gibbs energy). The mobility function is given by

\begin{equation} \label{eq:SubstrateAnchoring}
\mathbf{Q}(h,\phi) = \left[  \lambda \mathbf{I}  + \nu \left(
	\begin{array}{cc} 
		\cos 2\phi & \sin 2\phi \\
		- \sin 2\phi & \cos 2\phi
	\end{array} \right) \right] h^3,
\end{equation}
where $\mathbf{I}$ is the identity  matrix,
\begin{equation} \label{eq:AnisotropicViscosities}
	\lambda=\frac{2 + \eta}{4\left(1 + \eta\right)} \;, \quad
	\nu=-\frac{\eta}{4\left(1 + \eta\right)} \;, \quad
	\eta = \alpha_3 + \alpha_6\;,
\end{equation}
are anisotropic viscosities, and we will refer to $\eta$ as the anchoring anisotropy parameter. Note that for all NLCs for which we have data, $\eta\in (-1,0)$; we assume this henceforth.
In the special case $\eta=0$ (studied in our previous work\cite{Lam2018}) the governing equation no longer depends on the azimuthal director angle, $\phi$; this case is known as {\it degenerate} planar substrate anchoring. The Gibbs energy for our \gls{NLC} system is given by 
\begin{equation} \label{eq:FreeEnergy}
 E = \cC \left( 1 + \frac{\nabla h \cdot \nabla h }{2} \right) + \Psi(h) \;, 
\end{equation}
where the first term on the right hand side is the surface tension contribution; and the second term,
\begin{equation} \label{eq:DisjPressNLC}
	\Psi = - \pad{\Pi}{h} \;, \quad
	\Pi(h) = \cK\left[ \left(\frac{b}{h}\right)^3-\left(\frac{b}{h}\right)^2\right] + \frac{\cN}{2}\left[\frac{m(h)}{h}\right]^2 \;, 
\end{equation}
is the contribution from the effective disjoining pressure $\Pi(h)$, the first part of which is the power-law form of the disjoining pressure commonly used in the literature (see the review of Craster \& Matar~\cite{cm_rmp09} for an in depth discussion) consisting of Born repulsion and the van der Waals force; and the second term is the elastic contribution due to the antagonistic anchoring conditions.   The nondimensional coefficients $\cC$, $\cK$, and $\cN$, are the ratios of surface tension forces, disjoining pressure forces, and elastic forces respectively, to the viscous forces. The values of parameters used in our model simulations are based on experiments~\cite{Cazabat2011,Herminghaus1998}, as discussed in some detail in our earlier work~\cite{Lam2018}, and are set to
\begin{equation} \label{eq:NLC_paras}
	\cC = 0.0857 \;, \quad
	\cK = 36.0 \; ,\quad
	\cN = 1.67 \; ,\quad
	\beta = 1 \; ,\quad
	b = 0.01 \; ,
\end{equation}
values that lead to an effective disjoining pressure $\Pi$ of the form shown in Figure~\ref{fig:DisjoiningPressure}. The parameter $\eta$ is the key to the influence of spatially-varying substrate anchoring in the model: in line with values for the widely-used NLCs MBBA ($\eta=-0.42$~\cite{kneppe_jcp_1982}) and 5CB ($\eta=-0.45$~\cite{skarp_mclc_1980}) we use values in the range $\eta\in [-0.5,0]$ in our simulations.

\section{Linear Stability Analysis} \label{sec:LSA}

To gain initial insight into the effects of substrate anchoring ($\phi$) and the anchoring anisotropy parameter $\eta$,  we first simplify the governing equation by assuming uniform planar substrate anchoring ($\phi$ is constant, while $\theta=\pi/2$ on $z=0$). Furthermore, note that the coordinate system may be rotated such that the $x$ axis is parallel to the (uniform) anchoring; therefore, with no loss of generality we may assume $\phi=0$. We begin by using \gls{LSA}  to understand the stability of such a flat film. We consider flat films with free surface perturbations either parallel or perpendicular to the anchoring direction, specifically
\begin{equation} \label{eq:LSASolutionForms}
     h(x,t) = H_0 \left[ 1 + \ep e^{ \omega_\parallel t + q_\parallel x \mathrm{i}} \right] \quad \textrm{or} \quad
     h(y,t) = H_0 \left[ 1 + \ep e^{ \omega_\perp t + q_\perp y \mathrm{i}} \right] \;,
\end{equation}
where $\ep\ll1$; $q$ and $\omega$ are the wavenumber and growth rate of the perturbations; and $\parallel$ and $\perp$ subscripts respectively denote quantities parallel ($x$-direction) and perpendicular ($y$-direction) to the substrate anchoring $\phi$.

Substituting eqn (\ref{eq:LSASolutionForms}) into eqn (\ref{eq:GovEqn}) with $\phi=0$, the general form (dropping subscript notation) of the dispersion relations is 
\begin{equation} \label{eq:DispersionRelation}
	\omega = - \sigma H_0^3 q^2 \left[ \cC  q^2 - \Pi'(H_0)  \right] \;.
 \end{equation}
 The $\sigma H_0^3$ factor here arises from the mobility function of eqn (\ref{eq:SubstrateAnchoring}) and is present also for Newtonian films (for which $\cN=0=\nu$). The term in square brackets is a result of the Gibbs energy (\ref{eq:FreeEnergy}), which determines the transition between linear stability ($\Pi'(H_0) < 0$) and instability ($\Pi'(H_0) > 0$) as a function of the initial film thickness (see Figure~\ref{fig:DisjoiningPressure} for the stability regimes as they relate to the effective disjoining pressure and film height). By computing the most unstable wavenumber,
\begin{equation} \label{eq:MaxUnstableMode}
	q_{\parallel,m} = q_{\perp,m} = q_m = \sqrt{\frac{\Pi'(H_0)}{2\cC}} \;,
\end{equation}
it may be seen that the Gibbs energy determines the lengthscale of instabilities. We fix the mean film thickness to
the linearly unstable value $H_0=0.2$, referring the reader to our previous work~\cite{Lam2018,lam_jcp_2019} for a detailed study of how stability properties depend on mean film thickness. 

\begin{figure}[t!]
	\centering
	\includegraphics[width=0.45\textwidth]{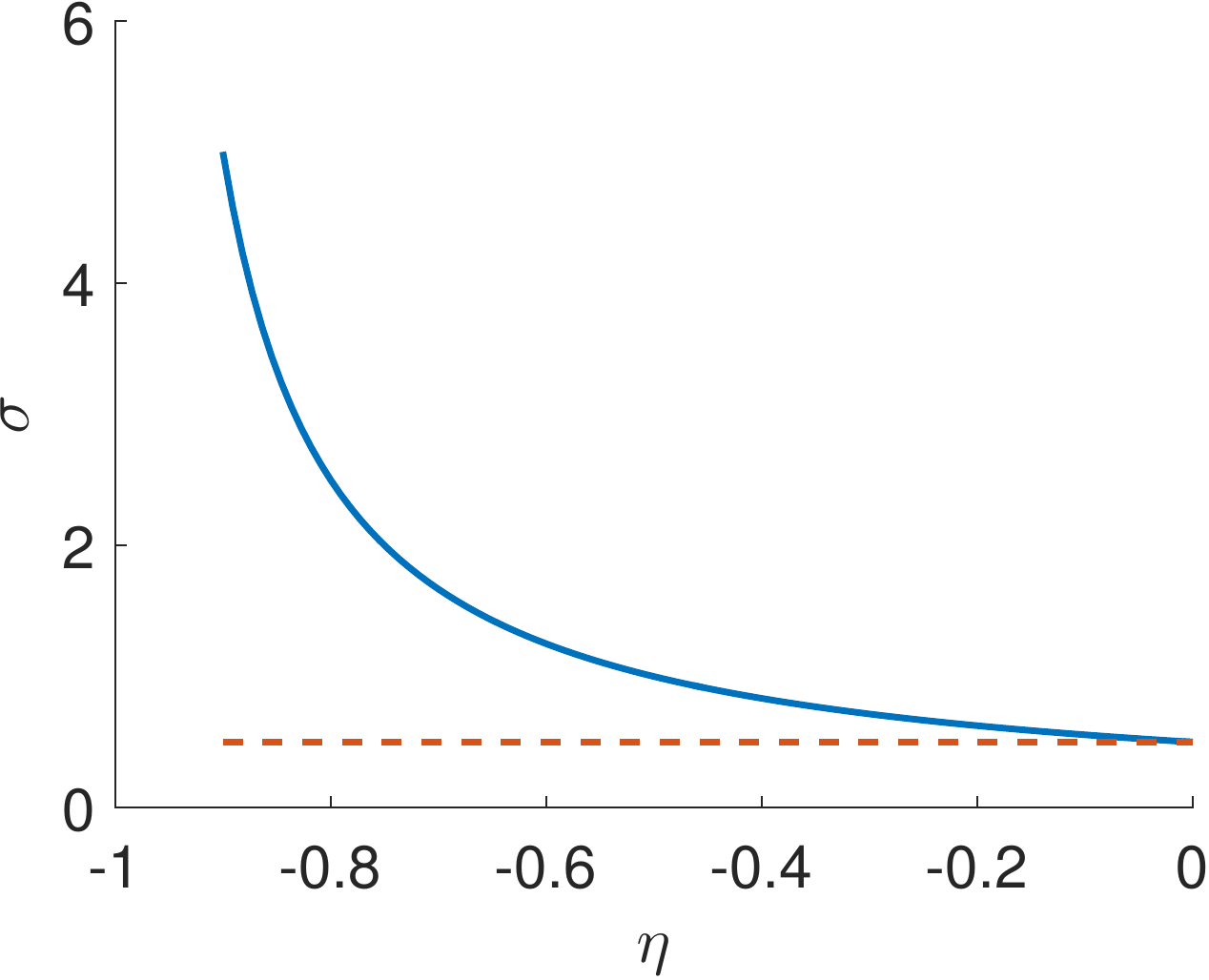} 
	\caption{ Plot of the scaling factors $\sigma_\parallel=\lambda+\nu=\left[2\left(1+\eta\right)\right]^{-1}$ (solid blue curve) and $\sigma_\perp=\lambda-\nu =0.5$ (dashed horizontal red line) as a function of $\eta\in(-1,0)$. }
	\label{fig:TimeScalingFactors}
\end{figure}

For a given value of $H_0$, the scaling factor $\sigma$ affects only the time-scale of instability. The most unstable growth rates are 
\begin{equation}  \label{eq:MaxUnstableGrowthrate}
	 \omega_{\parallel,m} = \sigma_\parallel \omega_m \; , \quad
	 \omega_{\perp,m} = \sigma_\perp \omega_m \; , \quad
	 \omega_m = \frac{H_0^3\left[\Pi'(H_0)\right]^2}{4\cC} \; .
\end{equation}
The scaling factors $\sigma_\parallel=\left[2 \left(1+\eta\right)\right]^{-1}$ (solid blue curve) and $\sigma_\perp=0.5$ (dashed horizontal red line) are plotted in Figure~\ref{fig:TimeScalingFactors} showing that, while perturbations in both $x$ and $y$ directions develop on the same timescale when $\eta=0$, the instability timescale increases with $|\eta|$ in the $x$-direction, while in the $y$-direction it is unchanged.

Considering the ratio of the growth rates
\begin{equation} \label{eq:GrowthrateRatio}
	 r = \frac{\omega_\parallel}{\omega_\perp} = \frac{\sigma_\parallel}{\sigma_\perp} = \frac{\lambda+\nu}{\lambda-\nu} = \frac{1}{1 + \eta} \;,
\end{equation}
it may be seen that for $\eta \in (-1,0)$ increasing $|\eta|$ increases the anisotropic viscous effects.
Similar to our previous works~\cite{Lam2018,lam_jcp_2019} for two-dimensional flow (where $\phi=0$ at the substrate), we define a new timescale,
\begin{equation} \label{eq:NewTimeScale}
 \tilde{t} = (\lambda+\nu)  t \;.
\end{equation}
Under the new timescale, the scaled growth rates are $\tilde\omega_{\parallel,m} =\omega_{m}$ and $\tilde\omega_{\perp,m} =(1+\eta)\omega_{m}$, and for $\eta < 0$, the maximum growth rate is independent of $\eta$. It is therefore expected that dewetting and drop formation occur on the time scale $\tilde{t}$ for all considered $\eta$ values and any observed differences in simulation results may likely be due to the influence of the value of $\eta$ on the nonlinear stages
of instability development.

To extend the LSA to a general three-dimensional film with height $h(x,y,t)$ (again in the special case $\phi=0$, strong substrate anchoring parallel to the $x$-direction), we first note that the governing equation (\ref{eq:GovEqn}) takes the form of a conservation law,

\begin{equation}
    \pad{h}{ t} + \nabla \cdot \mathbf{F}=0\;, \quad
	\mathbf{F} = \mathbf{Q} (h) \nabla \left( \frac{\delta E}{\delta h} \right) = \mathbf{Q} (h) \left[ \cC \nabla \nabla^2 h - \Pi'(h) \nabla h \right],\;
\end{equation}
with flux $\mathbf{F}$ ($E$ is defined in equation (\ref{eq:FreeEnergy})). 
Rescaling time as in (\ref{eq:NewTimeScale}), the governing equation may be expressed as 
\begin{equation} \label{eq:GovEqnReScaledTime}
	\pad{h}{\tilde t} + \nabla \cdot \tilde{\mathbf{F}}=0,
\end{equation}
where $\tilde{\mathbf{F}}=(\tilde F_x,\tilde F_y)=(\tilde F_\parallel,\tilde F_\perp)$, with
\begin{equation}
    \begin{split}
        \tilde F_\parallel & = \cC h^3 \partial_{xxx}  h - h^3\Pi'(h) \partial_{x} h + \cC h^3 \partial_{xyy} h \;, \quad {\rm and} \\
        \tilde F_\perp & = (1 + \eta) \left[ \cC h^3 \partial_{yyy}h  - h^3\Pi'(h) \partial_{y} h + \cC h^3 \partial_{yxx} h \right].
    \end{split}
\end{equation}
The limit $\eta \rightarrow - 1$ (or $|1+\eta| \ll 1$), corresponding to strong anisotropic effects, indicates 
that flux (flow) perpendicular to the substrate anchoring ($y$ direction) is negligible. Dependence on $y$ 
enters in this limit only via the last term in $\tilde F_\parallel$, which acts (via surface tension) to 
smooth any nonuniformities in the $y$-direction.

To apply LSA, we assume a solution of the form 
\begin{equation} \label{eq:LSASolutionForm2D}
     h(x,y,t) = H_0 \left[ 1 + \ep e^{ \omega t + q_\parallel x \mathrm{i} + q_\perp y \mathrm{i}} \right] ,
\end{equation}
with
\begin{equation} \label{eq:DispersionRelation2D}
	\omega = - H_0^3 \left[q_\parallel^2 + (1+\eta) q_\perp^2 \right] \left[ \cC  \left(q_\parallel^2 + q_\perp^2 \right) - \Pi'(H_0)  \right] \;.
 \end{equation}
Alternatively, if we express $(q_\parallel, q_\perp)$ in terms of a plane polar representation, i.e., $q_\parallel = q \cos (\vartheta)$ and $q_\perp =  q \sin (\vartheta)$, the dispersion relation equation (\ref{eq:DispersionRelation2D}) may be expressed as
\begin{equation} \label{eq:DispersionRelation2DPolar}
	\omega = - H_0^3 q^2 \left[ \cC q^2 - \Pi'(H_0) \right] \left( 1 + \frac{\eta}{2}  \left[ 1-\cos( 
2\vartheta)\right] \right) \;,
\end{equation}
an extension of the one-dimensional LSA dispersion relation equation (\ref{eq:DispersionRelation}) with 
$\sigma= 1 + \frac{\eta}{2}  \left[ 1-\cos(2\vartheta)\right]$. 

\section{Nonlinear regime}\label{sec:simulation}

In this section we present simulation results for the model outlined in the previous section. We focus in particular on the effects of imposed substrate anchoring patterns on dewetting, as revealed by the evolution of perturbed flat films.   We refer the reader to our previous works~\cite{Lam2018,lam_jcp_2019} for a detailed analysis of the stability properties of thin films of \gls{NLC} in the presence of degenerate planar substrate anchoring (as opposed to the non-degenerate, directed planar substrate anchoring considered here).   The numerical scheme is a hybrid finite volume/finite difference method with Crank-Nicholson type discretization. To reduce the computational complexity, an alternating direction implicit (ADI) scheme is implemented, and to account for implicit nonlinear terms a (psuedo)-Newton iterative scheme is applied. The reader is referred to our previous work for further details on the code~\cite{lam_jcp_2019}.

We divide our simulation results into three main categories: (1) uniform (unidirectional) anchoring (as discussed in Sec.~\ref{sec:LSA} above); (2) smoothly-varying nontrivial anchoring patterns; and (3) imposed substrate anchoring patterns that mimic defects in the director field.  To simulate the evolution of a randomly perturbed flat film of \gls{NLC}, perturbations are generated with pseudo-Perlin noise, sufficiently exciting all modes in the two-dimensional Fourier transform independently, see Lam {\it et al.}~\cite{lam_jcp_2019} for more details.  
Briefly, for all simulations in this paper, the initial condition is of the form
\begin{equation} \label{eq:3D_Random_IC}
 h(x,y,t=0) = H_0( 1 + \ep \left| \zeta(x,y) \right| ) \;, \quad
 \quad x,y\in[0, L] \;,
\end{equation}
where $H_0=0.2$, $L = P\lambda_m$, $P$ is a positive integer (set to 40 for all simulations in this paper), and $\lambda_m=2\pi/q_m$ is the wavelength of maximum growth, with $q_m$ given by equation (\ref{eq:MaxUnstableMode}).  The perturbations are specified by $\zeta(x,y)$, which is the inverse Fourier transform of 
\begin{equation}  \label{eq:PerlinNoise}
  \zeta(q_x,q_y)  =  \left| \left[ q_x^2 + q_y^2 \right]^{-\alpha/2} \exp \left(  2\pi a(q_x,q_y) \mathrm{i}  \right) \right|  \;,
\end{equation}
$\ep=0.01$ characterizes the noise amplitude, $\alpha$ is a positive constant, and $a(q_x,q_y)$ is a random variable, uniformly distributed on $[-1,1]$ for each $(q_x,q_y)$.   In addition, $\zeta(x,y)$ is scaled so that $|\zeta(x,y)|\leq1$ and we fix $\alpha=200/N$, where $N$  is the number of discretization points in the 
$x$ and $y$ directions.

\subsection{Uniform Substrate Anchoring} \label{sec:UniformAnchoring}

As noted earlier in Sec.~\ref{sec:LSA},  with uniform anchoring ($\phi$ constant) the coordinate system may be rotated appropriately, so that it is sufficient to consider the case $\phi=0$.   To study the effect of the anchoring anisotropy parameter $\eta $, which measures the strength of the directionality in the substrate anchoring via equations (\ref{eq:SubstrateAnchoring}) and (\ref{eq:AnisotropicViscosities}), we perform simulations for $\eta=0,-0.1,-0.2,-0.25,-0.35,$ and $-0.5$, which correspond to the growth rate ratios (defined by equation (\ref{eq:GrowthrateRatio})) $r\approx1, 1.1,1.25,1.33,1.5,$ and $2$, i.e. the growth rate of perturbations in the $x$ direction ranges from the same, up to twice as fast, as that in the $y$-direction.   In addition, to quantify the effect of the value of $\eta$ on the film evolution, we compute the magnitude of the two-dimensional Fourier transform of the perturbation to the free surface height.  To reduce the amount of noise, the Fourier transform is convoluted with a Gaussian filter~\cite{lam_jcp_2019}. 

\begin{figure}[t!]
	\centering
	\includegraphics[width=0.45\textwidth]{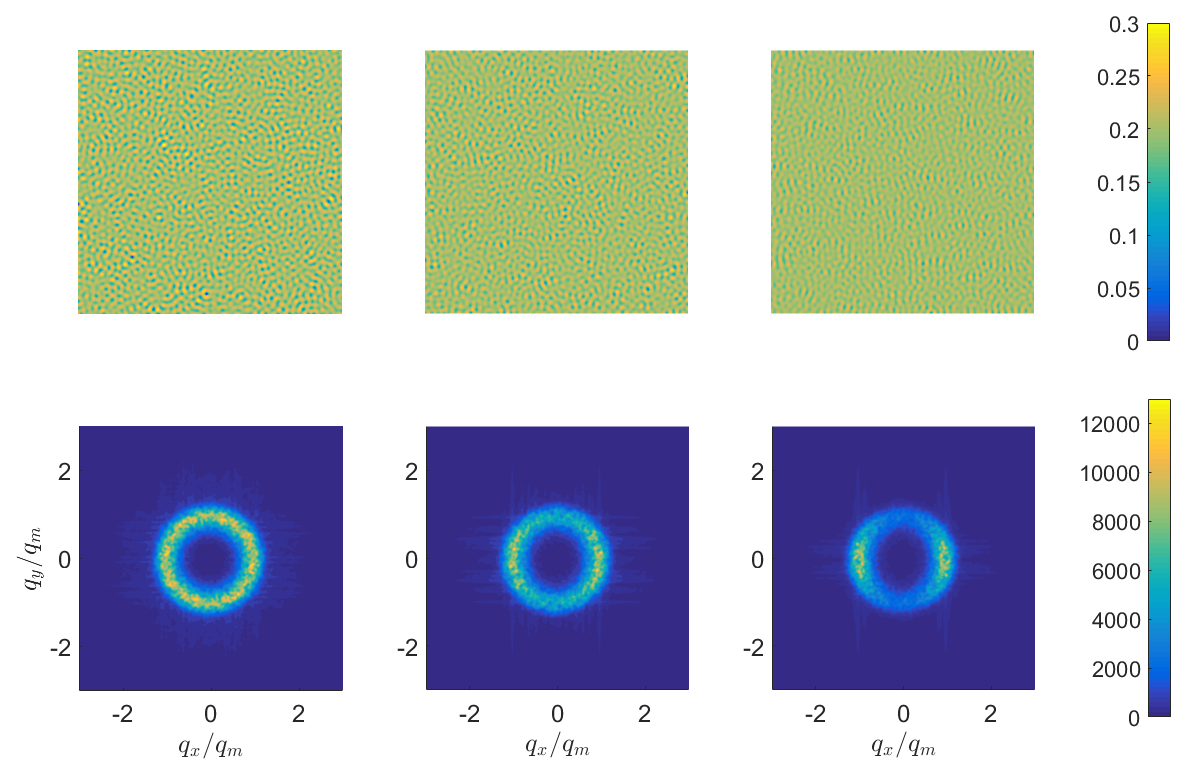} 
	\caption{ Snapshots of simulations with increasing absolute values (left to right) of the anchoring anisotropy parameter $\eta=0, -0.1, -0.2$, shown at scaled time $\tilde{t}=5 \omega_m^{-1}$ (see eqn (\ref{eq:NewTimeScale})). Top row shows contour plots of free surface height $h$, obtained by solving eqn (\ref{eq:GovEqn}), while bottom row shows the corresponding magnitude of the two-dimensional Fourier transform of the free surface height.
	For this figure (and all following ones) the domain size in both $x$ and $y$ directions 
	is $L = 40\lambda_m$, where $\lambda_m=2\pi/q_m$ is the wavelength of maximum growth, with $q_m$ given by eqn (\ref{eq:MaxUnstableMode}).
		}
	\label{fig:Eta_vary_tau_5_1}
\end{figure}

\begin{figure}[t!]
	\centering
	\includegraphics[width=0.45\textwidth]{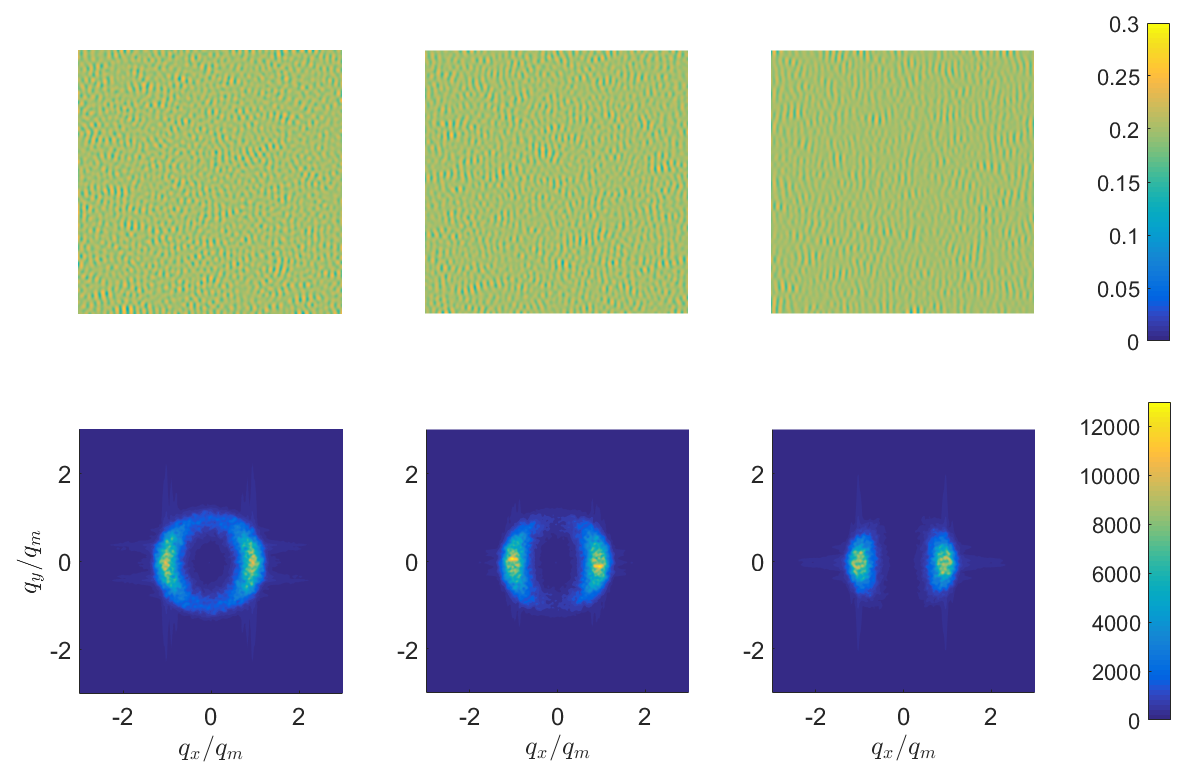} 
	\caption{Simulation results using (left to right) $\eta=-0.25,-0.35, -0.5$; the other parameters are as in Fig.~\ref{fig:Eta_vary_tau_5_1}. 
}
	\label{fig:Eta_vary_tau_5_2}
\end{figure}

\begin{figure}[t!]
	\centering
	\includegraphics[width=0.45\textwidth]{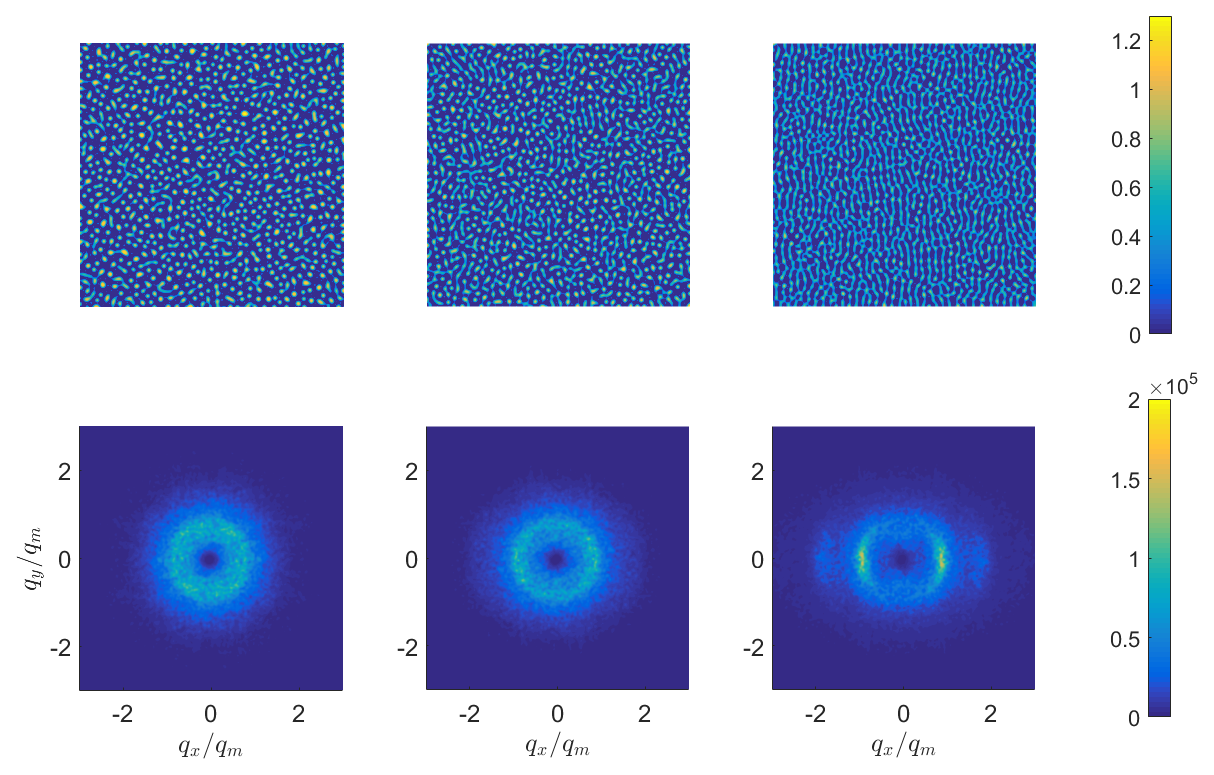} 
	\caption{Setup from Fig.~\ref{fig:Eta_vary_tau_5_1} at $\tilde{t}=8$.}
	\label{fig:Eta_vary_tau_8_1}
\end{figure}

\begin{figure}[t!]
	\centering
	\includegraphics[width=0.45\textwidth]{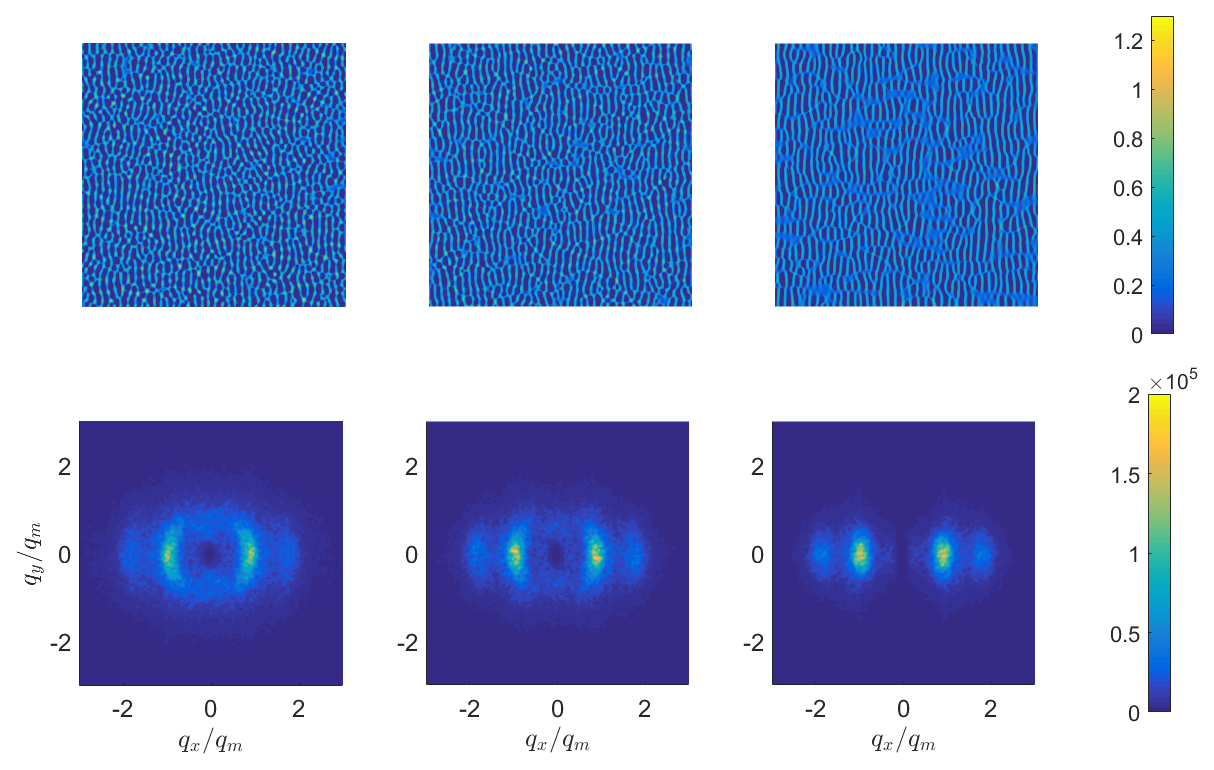} 
	\caption{Setup from Fig.~\ref{fig:Eta_vary_tau_5_2} at $\tilde{t}=8$.}
	\label{fig:Eta_vary_tau_8_2}
\end{figure}

\begin{figure}[t!]
	\centering
	\includegraphics[width=0.45\textwidth]{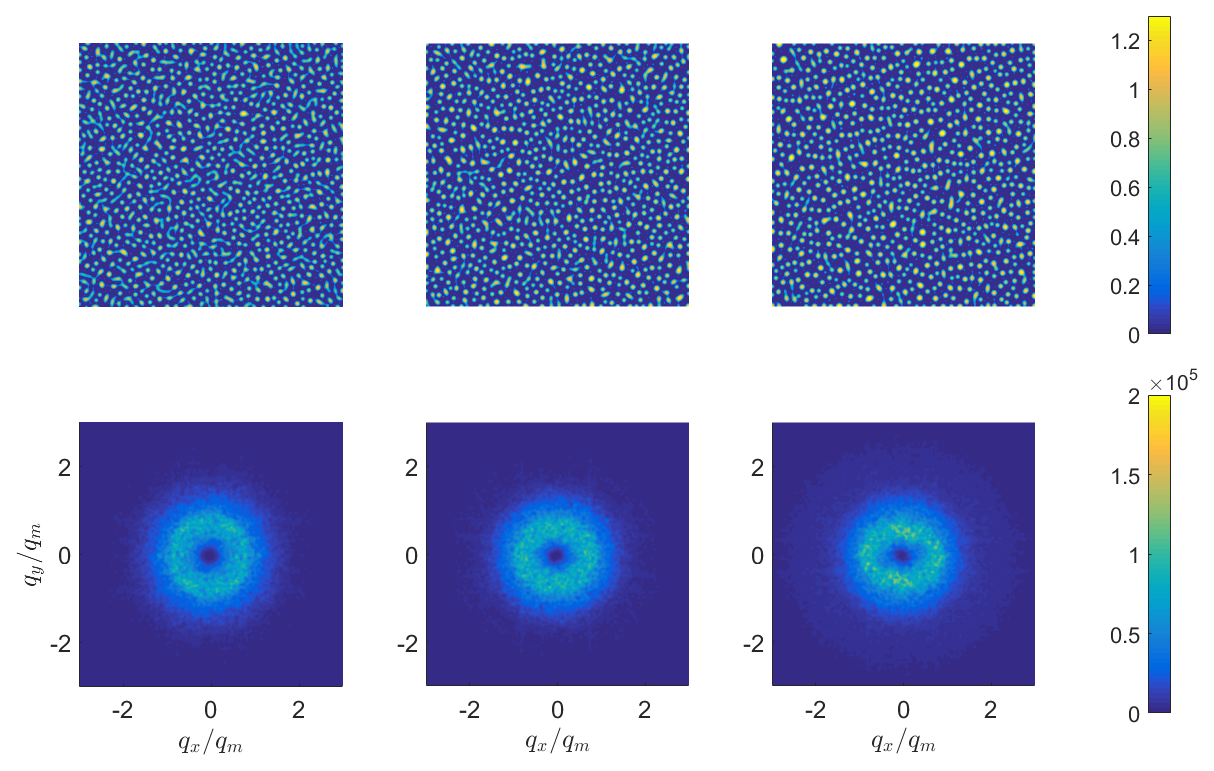} 
	\caption{Setup from Fig.~\ref{fig:Eta_vary_tau_5_1} at $\tilde{t}=500$.}
	\label{fig:Eta_vary_drops_1}
\end{figure}

\begin{figure}[t!]
	\centering
	\includegraphics[width=0.45\textwidth]{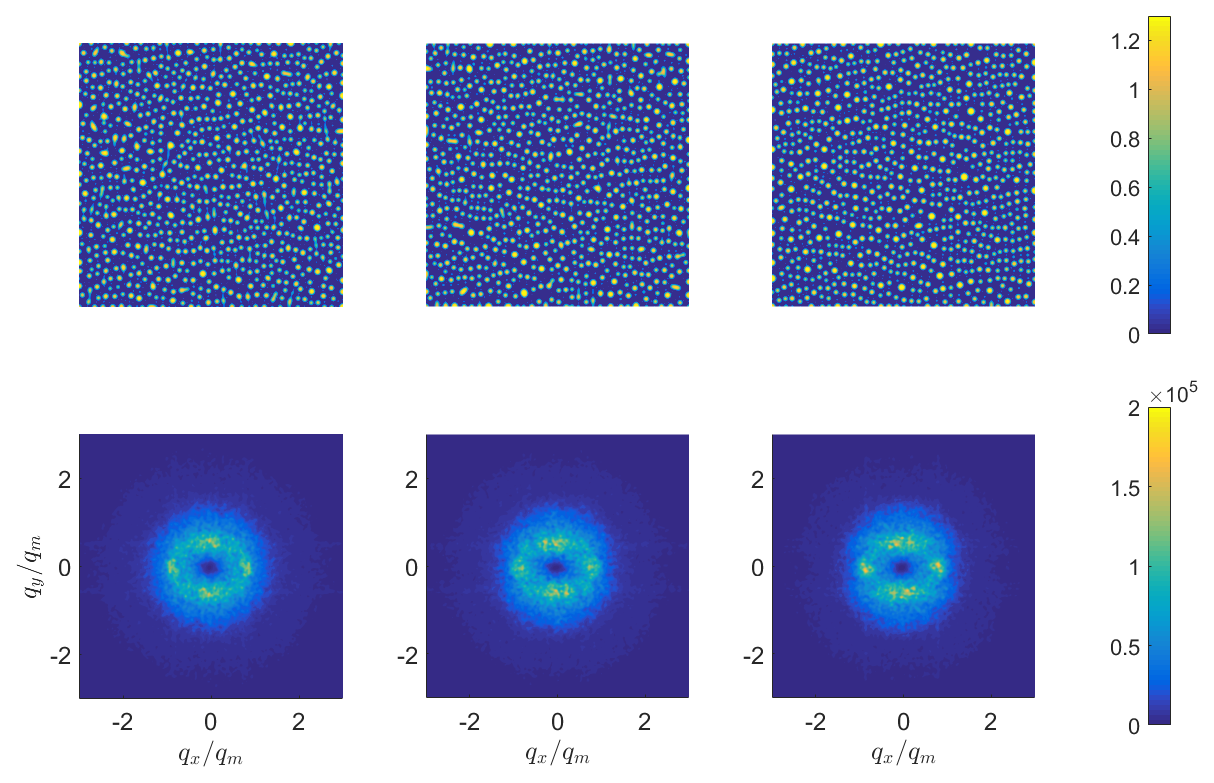} 
	\caption{ Setup from Fig.~\ref{fig:Eta_vary_tau_5_2} at $\tilde{t}=500$.
	}
	\label{fig:Eta_vary_drops_2}
\end{figure}

Figures~\ref{fig:Eta_vary_tau_5_1} and~\ref{fig:Eta_vary_tau_5_2} show early-time results for a flat film, perturbed as in equation (\ref{eq:3D_Random_IC}), evolving on a substrate with uniform anchoring in the $x$-direction.
(Later times are shown in Figures~\ref{fig:Eta_vary_tau_8_1} and \ref{fig:Eta_vary_tau_8_2}, and very late times in Figures~\ref{fig:Eta_vary_drops_1} and \ref{fig:Eta_vary_drops_2}, discussed below.)
The top rows show snapshots of the free surface evolution for the range of chosen $\eta$-values ($\eta=0, -0.1,$ and $-0.2$ for Figure~\ref{fig:Eta_vary_tau_5_1}, and $\eta=-0.25,-0.35,$ and $-0.5$ for Figure~\ref{fig:Eta_vary_tau_5_2}).  The free surface height in each case is plotted at the time $\tilde{t}=5\omega_m^{-1}$ (see equation (\ref{eq:NewTimeScale})), so that the instability should be equally well-developed for each case shown.  The simulations show that, as $|\eta|$ increases (left to right), the instability pattern becomes more wavelike, with ridges forming perpendicular to the anchoring direction. 

Focusing on the plots of the magnitude of the Fourier transform (bottom rows in Figures~\ref{fig:Eta_vary_tau_5_1} and~\ref{fig:Eta_vary_tau_5_2}), we observe that for degenerate planar substrate anchoring ($\eta=0$, bottom left part of Figure~\ref{fig:Eta_vary_tau_5_1}), the dominant wavenumbers form a ring, $(q_x/q_m)^2+(q_y/q_m)^2=1$; as expected, all wavenumbers that correspond to the most unstable wavenumber, $q_m$ from \gls{LSA} equation (\ref{eq:MaxUnstableMode}), are equally excited. Increasing $|\eta|$ concentrates the dominant wave numbers around $(q_x/q_m)^2=1$, indicating that perturbations along the $x$-axis dominate, confirming the \gls{LSA} prediction, see equation (\ref{eq:GrowthrateRatio}), for the ratio in growth rates: i.e.,  as $|\eta|$ increases, for the chosen uniform anchoring,  perturbations in the $x$-direction grow faster than in the $y$-direction.

It may also be seen by comparing Figs.~\ref{fig:Eta_vary_tau_5_1} and \ref{fig:Eta_vary_tau_5_2} (top row in each)
that for smaller values of $|\eta|$, more green/blue regions are observed in the height plots, indicating that dewetting occurs for early times. This conclusion is also supported by Figs.~\ref{fig:Eta_vary_tau_8_1} and
\ref{fig:Eta_vary_tau_8_2}, which show the same simulations as Figs.~\ref{fig:Eta_vary_tau_5_1}
and~\ref{fig:Eta_vary_tau_5_2}, but at the later time, $\tilde{t}=8$.   These figures show that, while drops have
formed for smaller values of $|\eta|$ (see Fig.~\ref{fig:Eta_vary_tau_8_1}), they have yet to do so for larger values
(see Fig.~~\ref{fig:Eta_vary_tau_8_2}).  This result is expected, as the maximum growth
rate in the $x$-direction is independent of $\eta$, while in the $y$-direction it decreases with increasing $|\eta|$.
Therefore, for small values of $|\eta|$, the drop formation is effectively a one-step process, since the instability 
develops on a similar timescale in both $x$ and $y$ directions.  However, for larger values of $|\eta|$, drop 
formation is a two-step process: first the ridges (approximately parallel to the $y$-axis) need to form due to the 
(fast) instability in the $x$ direction, and then these ridges need to break, a process that occurs on a slower timescale. 

To conclude this section, we show the simulations for the same parameters used 
in Figs.~\ref{fig:Eta_vary_tau_5_1}-\ref{fig:Eta_vary_tau_8_2} at a very late 
time $\tilde{t} = 500$ (chosen to be sufficiently large that the film has fully dewetted and no further evolution is anticipated). Figures~\ref{fig:Eta_vary_drops_1} and~\ref{fig:Eta_vary_drops_2}
show that by this stage complete film breakup into droplets has occurred in all cases. 
Although the drop sizes and spacing do not differ significantly as $|\eta|$ varies, 
for strong anchoring anisotropy we observe drops forming along aligned `tracks'. 
This drop alignment is due to the ridge formation that occurred prior to drop 
formation, and is confirmed by the slight anisotropy observed in the Fourier 
transform data (see in particular the bottom row of Figure~\ref{fig:Eta_vary_drops_2}). These very large time results confirm that the effect of directed substrate anchoring will always be evident in the system, even after complete dewetting.
 
\subsection{Spatially Continuous Substrate Anchoring}

Having gained some insight into the effect of the anchoring anisotropy parameter $\eta$, we now 
fix $\eta=-0.5$ for the remaining results and investigate the initial dewetting process (before the final drop formation) for more exotic substrate anchoring patterns. The value chosen for $\eta$ here is representative of typical values for common NLCs, motivated by the values reported for two widely-used NLCs MBBA ($\eta=-0.42$~\cite{kneppe_jcp_1982}) and 5CB ($\eta=-0.45$~\cite{skarp_mclc_1980}). We first consider a continuous choice for the imposed azimuthal {substrate anchoring pattern, the so-called ``egg carton'' potential, $\phi_{\rm S}(x,y)$: 
\begin{equation} \label{eq:ContinousPattern}
	\phi_{\rm S}(x,y) = \frac{\pi}{2} \cos\left( \frac{ 8 x}{L}\right) \cos\left( \frac{ 4 y}{L} \right) \;, \quad 
    x,y\in[0, L] \;, 
\end{equation}
(recall the polar angle $\theta$ is fixed at $\pi/2$ on the substrate due to the strong planar anchoring) where as before, $L=P\lambda_m$ with $P=40$.

Figure~\ref{fig:continuous_pattern_1} shows snapshots of the evolving free surface height at two 
time instances, before dewetting and as dewetting is occurring, as in Figures~\ref{fig:Eta_vary_tau_5_1} and~\ref{fig:Eta_vary_tau_8_1}, respectively. The anchoring pattern specified by equation (\ref{eq:ContinousPattern}) is overlaid in white.  This simulation demonstrates that the general predictions from the \gls{LSA} results for uniform anchoring extend to this more complicated anchoring scenario: ridge formation occurs first perpendicular to the local anchoring direction. At later times (not shown here) the ridges themselves undergo breakup along their length, into droplets. As with the uniform substrate anchoring case, the final droplet configuration obtained at large times retains the characteristics of the underlying anchoring pattern, with droplets aligned along ``tracks'' where the initial ridges formed.

\begin{figure}
	\centering
	\includegraphics[width=0.225\textwidth]{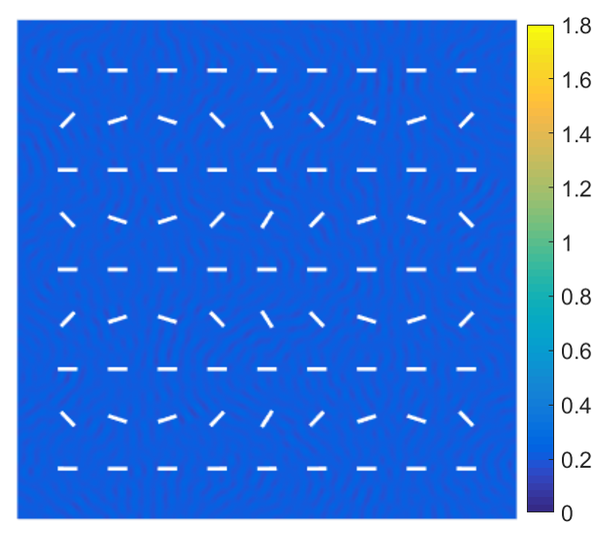} 
	\includegraphics[width=0.225\textwidth]{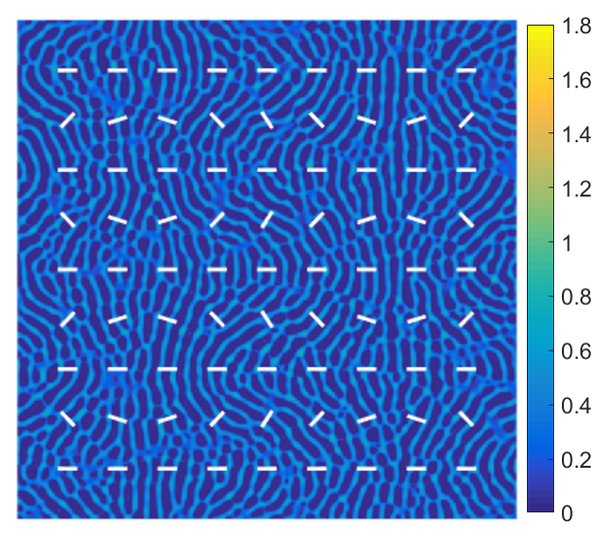} 
	\caption{Plot of the evolving free surface height at $\tilde{t}=5$ (left) and $\tilde{t}=8$ (right) with continuous spatially-varying substrate anchoring (overlaid in white) given by eqn (\ref{eq:ContinousPattern}). Anchoring anisotropy parameter $\eta=-0.5$.}
	\label{fig:continuous_pattern_1}
\end{figure}

\subsection{Anchoring with ``Defect'' Patterns}
To conclude our numerical study, we present selected simulations of film evolution over substrates with anchoring patterns that contain imposed topological defects at the origin; specifically, $\phi$ is of the form
\begin{equation} \label{eq:DefectPattern}
	\phi_{\rm D}(x,y;s) = s \, \textrm{arctan}(x/y)  \;, \quad
	x,y\in\left[-\frac{L}{2}, \frac{L}{2} \right] \;, 
\end{equation}
where $s$ is the topological winding number of the defect, measuring the number of rotations of the director angle $\phi$ as a small planar circuit is traversed anticlockwise around the defect (i.e., traversing such a circuit, $\phi$ changes from $0$ to $ 2\pi s$). Note that at the defect location the underlying Leslie-Ericksen model breaks down (an accurate description of defects in NLCs requires a more sophisticated model, such as Landau-De Gennes, with an order parameter that allows for localized ``melting" of the director structure at the defect core); nonetheless, we expect that our model equation (\ref{eq:GovEqn}) may provide a reasonable qualitative indication of how the presence of a defect influences the overlying flow, in particular, the film thickness evolution.

In Figures~\ref{fig:Defect_Single_tau_5} and~\ref{fig:Defect_Single_tau_8}, the free surface height is plotted at the same dimensionless scaled times as before, $\tilde{t}=5$ and $\tilde{t}=8$ respectively, for four different values of the topological winding number $s$: (a) $s=1/2$; (b) $s=-1$; (c) $s=-1/2$; and (d) $s=1$.   
The anchoring patterns are again overlaid on the free surface contour plot to emphasise how ridges form in the surface perpendicular to the anchoring pattern.   

\begin{figure}
	\centering
	\subfigure[]{
	\includegraphics[width=0.225\textwidth]{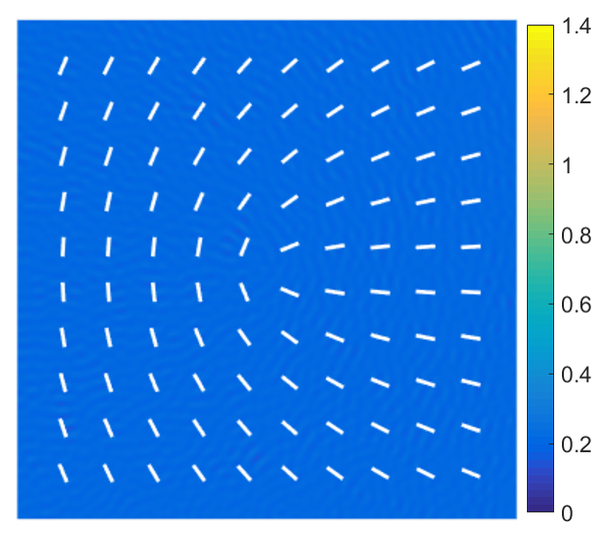} 
	\label{fig:Defect_Single_Type_4_tau_5} }
	\subfigure[]{
	\includegraphics[width=0.225\textwidth]{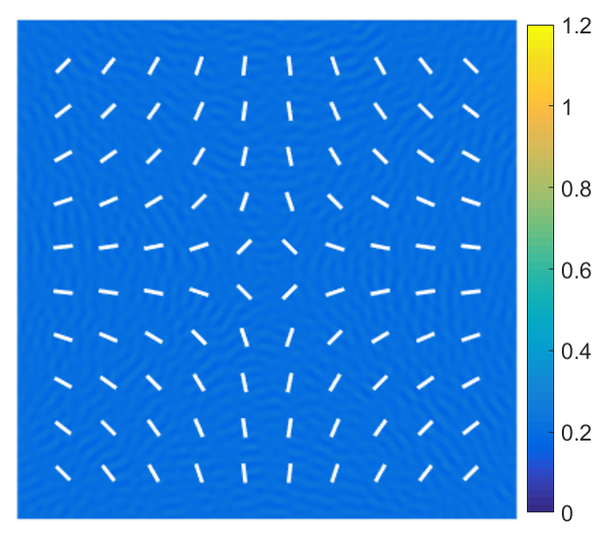}
	\label{fig:Defect_Single_Type_1_tau_5} }
	\subfigure[]{
	\includegraphics[width=0.225\textwidth]{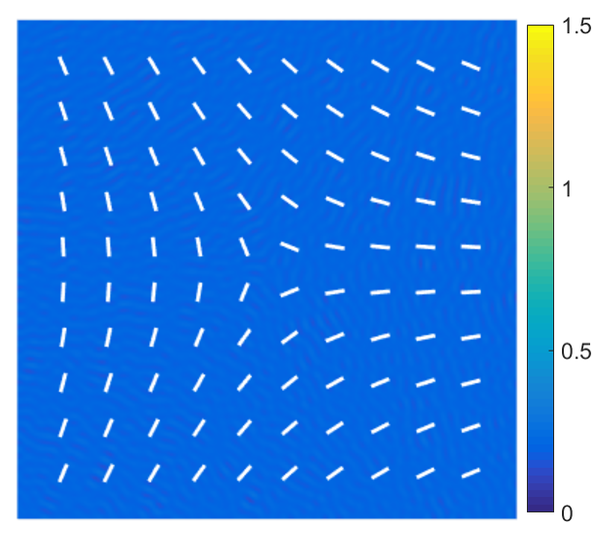} 
	\label{fig:Defect_Single_Type_3_tau_5} }
	\subfigure[]{
	\includegraphics[width=0.225\textwidth]{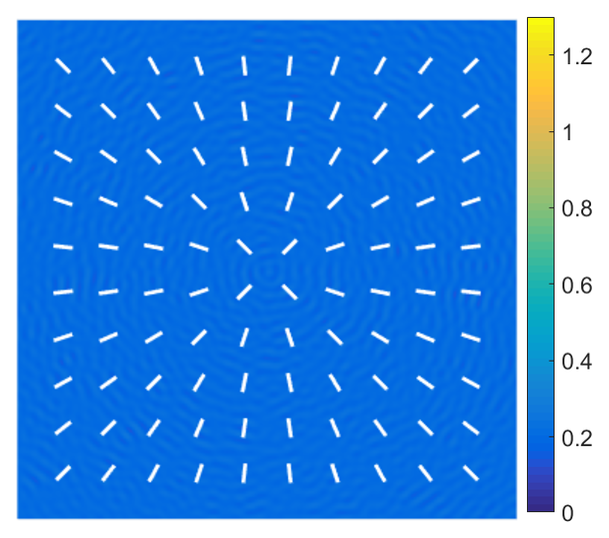}
	\label{fig:Defect_Single_Type_2_tau_5} }
	\caption{Plot of the evolving free surface height at $\tilde{t}=5$ for defect anchoring patterns (overlaid in white) defined by eqn (\ref{eq:DefectPattern}) with winding numbers (a) $s=1/2$, (b) $s=-1$, (c) $s=-1/2$, and (d) $s=1$. Anchoring anisotropy parameter $\eta=-0.5$.}
	\label{fig:Defect_Single_tau_5}
\end{figure}

\begin{figure}
	\centering
	\subfigure[]{
	\includegraphics[width=0.225\textwidth]{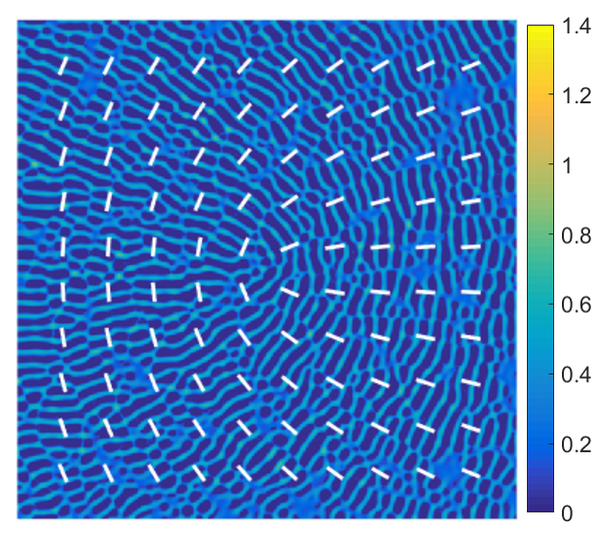} 
	\label{fig:Defect_Single_Type_4_tau_8} }
	\subfigure[]{
	\includegraphics[width=0.225\textwidth]{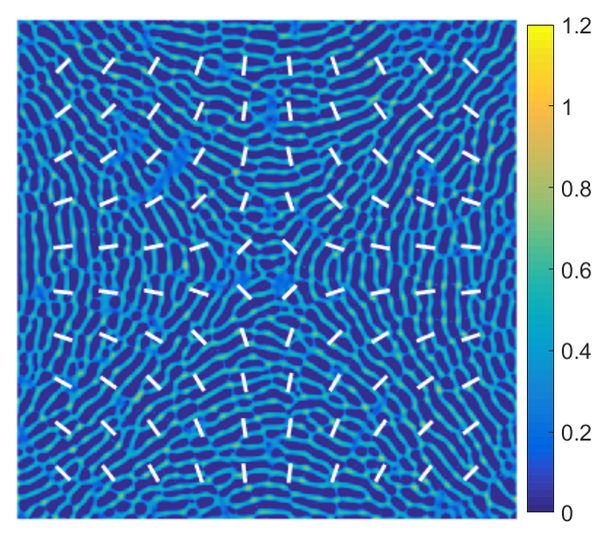}
	\label{fig:Defect_Single_Type_1_tau_8} }
	\subfigure[]{
	\includegraphics[width=0.225\textwidth]{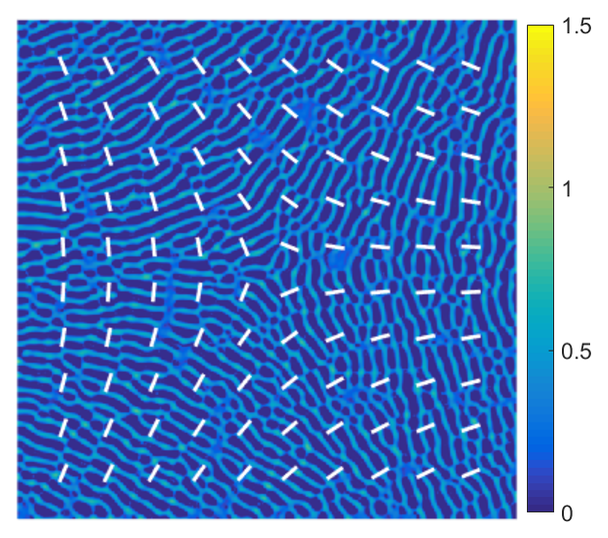} 
	\label{fig:Defect_Single_Type_3_tau_8} }
	\subfigure[]{
	\includegraphics[width=0.225\textwidth]{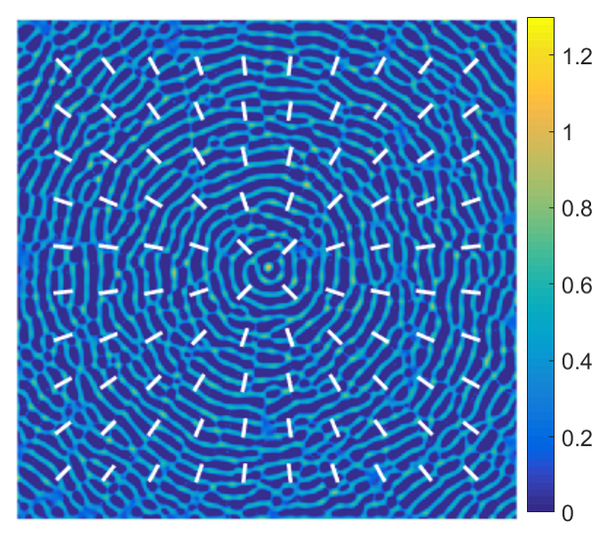}
	\label{fig:Defect_Single_Type_2_tau_8} }
	\caption{Setup from Fig.~\ref{fig:Defect_Single_tau_5} at $\tilde{t} = 8$.}
	\label{fig:Defect_Single_tau_8}
\end{figure}
   
To complete this section, we present a simulation of flow over an anchoring pattern that incorporates multiple such idealized defects.   To construct a multi-defect pattern, the domain is divided into quadrants, each containing one defect, and hyperbolic tangent functions are used to connect neighboring quadrants smoothly.  Figure~\ref{fig:Defect_Multi_Type_1} plots the free surface height at $\tilde{t}=5$ (a) and  $\tilde{t}=8$ (b)
for a simulation with an anchoring pattern containing all four of the defects from  Fig.~\ref{fig:Defect_Single_tau_5}. Similarly to our previous results, initial dewetting leads to rivulets, which very clearly form perpendicular to the underlying substrate anchoring pattern. Ultimately (not shown here 
for brevity), these rivulets undergo breakup into droplets, the arrangement of which reflects the 
imposed substrate pattern.
   
\begin{figure}
	\centering
		\subfigure[$\tilde{t}=5$]{
	\includegraphics[width=0.475\textwidth]{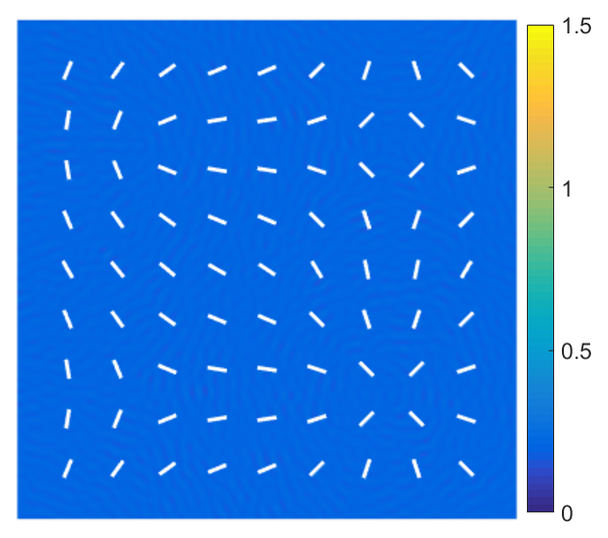} }
		\subfigure[$\tilde{t}=8$]{
	\includegraphics[width=0.475\textwidth]{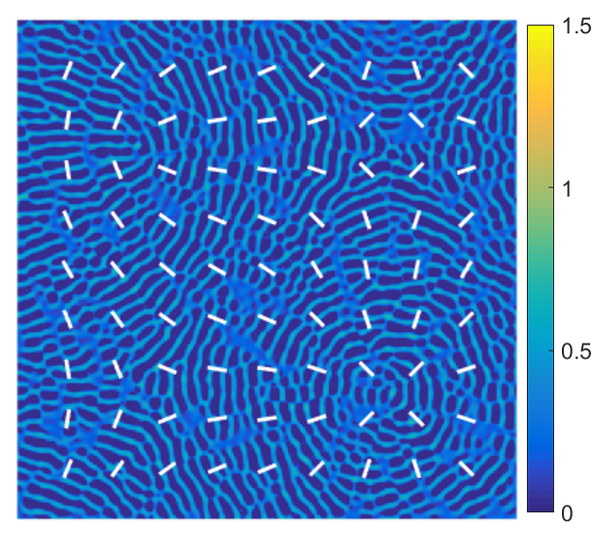} }
	\caption{Free surface height evolution over a substrate 
	with variable anchoring incorporating all four considered defect types.}
	\label{fig:Defect_Multi_Type_1}
\end{figure}

\section{Conclusions}\label{sec:conclusions}

We have presented a simple model that allows us to explore the effects of inhomogeneous planar substrate
anchoring on the free surface evolution of dewetting thin \gls{NLC} films. 
Multiple simulations show that 
with sufficiently large anisotropic viscous effects, represented by larger absolute values of the (negative) 
parameter $\eta$ (see equation (\ref{eq:AnisotropicViscosities})), the intermediate-time dynamics of the film are strongly affected relative to the degenerate planar anchoring case, with ridges, and later rivulets, forming
perpendicular to the imposed local anchoring direction, a result consistent with the predictions of linear stability analysis (\gls{LSA}).   Furthermore, the underlying characteristic lengthscale of instability, given by 
$2\pi/q_m$ (see equation (\ref{eq:MaxUnstableMode})), is unaffected by the value of the anchoring anisotropy 
parameter $\eta$. At very long times our numerical results show that $\eta$ has a clear effect on the final spatial distribution of drops. 

In addition to our simulations on substrates with imposed smoothly-varying anchoring patterns, we also present a number of simulations of thin film flow over idealized substrate ``defects''.   Though such structures clearly influence the morphology of the evolving film, they do not (within the limitations of the model presented here) themselves destabilize the film, as might be expected. It would, therefore, be of interest in future to consider physically-motivated modifications to the governing equation (based, {\it e.g.} on the Landau-DeGennes framework, which allows for a more realistic treatment of defects), to determine with greater certainty whether defects may be capable of inducing instability.

\section*{Conflicts of interest}
There are no conflicts to declare.

\subsubsection*{Acknowledgements}
All authors acknowledge support from the National Science Foundation under NSF DMS 1211713 and NSF DMS 1815613.

\balance


\bibliography{films.bib}

\providecommand*{\mcitethebibliography}{\thebibliography}
\csname @ifundefined\endcsname{endmcitethebibliography}
{\let\endmcitethebibliography\endthebibliography}{}
\begin{mcitethebibliography}{30}
\providecommand*{\natexlab}[1]{#1}
\providecommand*{\mciteSetBstSublistMode}[1]{}
\providecommand*{\mciteSetBstMaxWidthForm}[2]{}
\providecommand*{\mciteBstWouldAddEndPuncttrue}
  {\def\EndOfBibitem{\unskip.}}
\providecommand*{\mciteBstWouldAddEndPunctfalse}
  {\let\EndOfBibitem\relax}
\providecommand*{\mciteSetBstMidEndSepPunct}[3]{}
\providecommand*{\mciteSetBstSublistLabelBeginEnd}[3]{}
\providecommand*{\EndOfBibitem}{}
\mciteSetBstSublistMode{f}
\mciteSetBstMaxWidthForm{subitem}
{(\emph{\alph{mcitesubitemcount}})}
\mciteSetBstSublistLabelBeginEnd{\mcitemaxwidthsubitemform\space}
{\relax}{\relax}

\bibitem[Marchetti \emph{et~al.}(2013)Marchetti, Joanny, Ramaswamy, Liverpool,
  Prost, Rao, and Simha]{marchetti_rmp_2013}
M.~C. Marchetti, J.~F. Joanny, S.~Ramaswamy, T.~B. Liverpool, J.~Prost, M.~Rao
  and R.~A. Simha, \emph{Rev. Mod. Phys.}, 2013, \textbf{85}, 1143--1189\relax
\mciteBstWouldAddEndPuncttrue
\mciteSetBstMidEndSepPunct{\mcitedefaultmidpunct}
{\mcitedefaultendpunct}{\mcitedefaultseppunct}\relax
\EndOfBibitem
\bibitem[Saintillan(2018)]{saintillan_arfm_2018}
D.~Saintillan, \emph{Annual Rev. Fluid Mech.}, 2018, \textbf{50},
  563--592\relax
\mciteBstWouldAddEndPuncttrue
\mciteSetBstMidEndSepPunct{\mcitedefaultmidpunct}
{\mcitedefaultendpunct}{\mcitedefaultseppunct}\relax
\EndOfBibitem
\bibitem[Sankararaman and Ramaswamy(2009)]{ramaswamy_prl_2009}
S.~Sankararaman and S.~Ramaswamy, \emph{Phys. Rev. Lett.}, 2009, \textbf{102},
  118107\relax
\mciteBstWouldAddEndPuncttrue
\mciteSetBstMidEndSepPunct{\mcitedefaultmidpunct}
{\mcitedefaultendpunct}{\mcitedefaultseppunct}\relax
\EndOfBibitem
\bibitem[Joanny and Ramaswamy(2012)]{joanny_jfm_2012}
J.-F. Joanny and S.~Ramaswamy, \emph{J. Fluid Mech.}, 2012, \textbf{705},
  46--57\relax
\mciteBstWouldAddEndPuncttrue
\mciteSetBstMidEndSepPunct{\mcitedefaultmidpunct}
{\mcitedefaultendpunct}{\mcitedefaultseppunct}\relax
\EndOfBibitem
\bibitem[Trinschek \emph{et~al.}(2020)Trinschek, Stegemerten, John, and
  Thiele]{thiele_pre_2020}
S.~Trinschek, F.~Stegemerten, K.~John and U.~Thiele, \emph{Phys. Rev. E}, 2020,
  \textbf{101}, 062802\relax
\mciteBstWouldAddEndPuncttrue
\mciteSetBstMidEndSepPunct{\mcitedefaultmidpunct}
{\mcitedefaultendpunct}{\mcitedefaultseppunct}\relax
\EndOfBibitem
\bibitem[Blow \emph{et~al.}(2017)Blow, Aqil, Liebchen, and
  Marenduzzo]{blow_sm_2017}
M.~L. Blow, M.~Aqil, B.~Liebchen and D.~Marenduzzo, \emph{Soft Matter}, 2017,
  \textbf{13}, 6137--6144\relax
\mciteBstWouldAddEndPuncttrue
\mciteSetBstMidEndSepPunct{\mcitedefaultmidpunct}
{\mcitedefaultendpunct}{\mcitedefaultseppunct}\relax
\EndOfBibitem
\bibitem[Cazabat \emph{et~al.}(2011)Cazabat, Delabre, Richard, Sang, and
  C.]{Cazabat2011}
A.~M. Cazabat, U.~Delabre, C.~Richard, Y.~Sang and Y.~C., \emph{Adv. Colloid
  Interface Sci.}, 2011, \textbf{168}, 29\relax
\mciteBstWouldAddEndPuncttrue
\mciteSetBstMidEndSepPunct{\mcitedefaultmidpunct}
{\mcitedefaultendpunct}{\mcitedefaultseppunct}\relax
\EndOfBibitem
\bibitem[Delabre \emph{et~al.}(2009)Delabre, Richard, and Cazabat]{Delabre2009}
U.~Delabre, C.~Richard and A.~M. Cazabat, \emph{J. Phys. Chem. B}, 2009,
  \textbf{113}, 3647\relax
\mciteBstWouldAddEndPuncttrue
\mciteSetBstMidEndSepPunct{\mcitedefaultmidpunct}
{\mcitedefaultendpunct}{\mcitedefaultseppunct}\relax
\EndOfBibitem
\bibitem[Herminghaus \emph{et~al.}(1998)Herminghaus, Jacobs, Mecke, Bischof,
  Fery, Ibn-Elhaj, and Schlagowski]{Herminghaus1998}
S.~Herminghaus, K.~Jacobs, K.~Mecke, J.~Bischof, A.~Fery, M.~Ibn-Elhaj and
  S.~Schlagowski, \emph{Science}, 1998, \textbf{282}, 916\relax
\mciteBstWouldAddEndPuncttrue
\mciteSetBstMidEndSepPunct{\mcitedefaultmidpunct}
{\mcitedefaultendpunct}{\mcitedefaultseppunct}\relax
\EndOfBibitem
\bibitem[van Effenterre and Valignat(2003)]{Effenterre2003}
D.~van Effenterre and M.~P. Valignat, \emph{Eur. Phys. J. E}, 2003,
  \textbf{12}, 367\relax
\mciteBstWouldAddEndPuncttrue
\mciteSetBstMidEndSepPunct{\mcitedefaultmidpunct}
{\mcitedefaultendpunct}{\mcitedefaultseppunct}\relax
\EndOfBibitem
\bibitem[Lam \emph{et~al.}(2018)Lam, Cummings, and Kondic]{Lam2018}
M.-A. Y.-H. Lam, L.~Cummings and Kondic, \emph{J Fluid Mech.}, 2018,
  \textbf{841}, 925\relax
\mciteBstWouldAddEndPuncttrue
\mciteSetBstMidEndSepPunct{\mcitedefaultmidpunct}
{\mcitedefaultendpunct}{\mcitedefaultseppunct}\relax
\EndOfBibitem
\bibitem[Lam \emph{et~al.}(2018)Lam, Cummings, and Kondic]{lam_jcp_2019}
M.-A. Y.-H. Lam, L.~Cummings and Kondic, \emph{J Comput. Phys. X}, 2018,
  \textbf{2}, 100001\relax
\mciteBstWouldAddEndPuncttrue
\mciteSetBstMidEndSepPunct{\mcitedefaultmidpunct}
{\mcitedefaultendpunct}{\mcitedefaultseppunct}\relax
\EndOfBibitem
\bibitem[Kim \emph{et~al.}(2001)Kim, Yoneya, Yamamoto, and
  Yokoyama]{apl_yokoyama2001}
J.-H. Kim, M.~Yoneya, J.~Yamamoto and H.~Yokoyama, \emph{Appl. Phys. Lett.},
  2001, \textbf{78}, 3055\relax
\mciteBstWouldAddEndPuncttrue
\mciteSetBstMidEndSepPunct{\mcitedefaultmidpunct}
{\mcitedefaultendpunct}{\mcitedefaultseppunct}\relax
\EndOfBibitem
\bibitem[Greschek \emph{et~al.}(2012)Greschek, Gubbins, and
  Shoen]{jcp_schoen2012}
M.~Greschek, K.~Gubbins and M.~Shoen, \emph{J. Chem. Phys.}, 2012,
  \textbf{137}, 144703\relax
\mciteBstWouldAddEndPuncttrue
\mciteSetBstMidEndSepPunct{\mcitedefaultmidpunct}
{\mcitedefaultendpunct}{\mcitedefaultseppunct}\relax
\EndOfBibitem
\bibitem[Kim \emph{et~al.}(2002)Kim, Yoneya, Yamamoto, and
  Yokoyama]{nature_yokoyama2002}
J.-H. Kim, M.~Yoneya, J.~Yamamoto and H.~Yokoyama, \emph{Nature}, 2002,
  \textbf{420}, 159\relax
\mciteBstWouldAddEndPuncttrue
\mciteSetBstMidEndSepPunct{\mcitedefaultmidpunct}
{\mcitedefaultendpunct}{\mcitedefaultseppunct}\relax
\EndOfBibitem
\bibitem[Forest \emph{et~al.}(2012)Forest, Wang, and Yang]{forest_sm_2012}
M.~G. Forest, Q.~Wang and X.~Yang, \emph{Soft Matter}, 2012, \textbf{8},
  9642--9660\relax
\mciteBstWouldAddEndPuncttrue
\mciteSetBstMidEndSepPunct{\mcitedefaultmidpunct}
{\mcitedefaultendpunct}{\mcitedefaultseppunct}\relax
\EndOfBibitem
\bibitem[Lin \emph{et~al.}(2013)Lin, Kondic, Thiele, and Cummings]{Lin2013}
T.-S. Lin, L.~Kondic, U.~Thiele and L.~J. Cummings, \emph{J. Fluid Mech.},
  2013, \textbf{729}, 214\relax
\mciteBstWouldAddEndPuncttrue
\mciteSetBstMidEndSepPunct{\mcitedefaultmidpunct}
{\mcitedefaultendpunct}{\mcitedefaultseppunct}\relax
\EndOfBibitem
\bibitem[Rey(2007)]{rey_sm_2007}
A.~D. Rey, \emph{Soft Matter}, 2007, \textbf{3}, 1349--1368\relax
\mciteBstWouldAddEndPuncttrue
\mciteSetBstMidEndSepPunct{\mcitedefaultmidpunct}
{\mcitedefaultendpunct}{\mcitedefaultseppunct}\relax
\EndOfBibitem
\bibitem[Rey and Herrera-Valencia(2014)]{rey_sm_2010}
A.~D. Rey and E.~Herrera-Valencia, \emph{Soft Matter}, 2014, \textbf{10},
  1611--1620\relax
\mciteBstWouldAddEndPuncttrue
\mciteSetBstMidEndSepPunct{\mcitedefaultmidpunct}
{\mcitedefaultendpunct}{\mcitedefaultseppunct}\relax
\EndOfBibitem
\bibitem[Leslie(1979)]{Leslie1979}
F.~M. Leslie, \emph{Adv. Liq. Cryst.}, 1979, \textbf{4}, 1\relax
\mciteBstWouldAddEndPuncttrue
\mciteSetBstMidEndSepPunct{\mcitedefaultmidpunct}
{\mcitedefaultendpunct}{\mcitedefaultseppunct}\relax
\EndOfBibitem
\bibitem[Rey(2008)]{Rey2008}
A.~D. Rey, \emph{Mol. Cryst. Liq. Cryst.}, 2008, \textbf{369}, 63\relax
\mciteBstWouldAddEndPuncttrue
\mciteSetBstMidEndSepPunct{\mcitedefaultmidpunct}
{\mcitedefaultendpunct}{\mcitedefaultseppunct}\relax
\EndOfBibitem
\bibitem[Ziherl and \u{Z}umer(2003)]{Ziherl2003}
P.~Ziherl and S.~\u{Z}umer, \emph{Eur. Phys. J. E}, 2003, \textbf{12},
  361\relax
\mciteBstWouldAddEndPuncttrue
\mciteSetBstMidEndSepPunct{\mcitedefaultmidpunct}
{\mcitedefaultendpunct}{\mcitedefaultseppunct}\relax
\EndOfBibitem
\bibitem[Cummings(2004)]{Cummings2004}
L.~J. Cummings, \emph{Eur. J. Appl. Maths}, 2004, \textbf{15}, 651\relax
\mciteBstWouldAddEndPuncttrue
\mciteSetBstMidEndSepPunct{\mcitedefaultmidpunct}
{\mcitedefaultendpunct}{\mcitedefaultseppunct}\relax
\EndOfBibitem
\bibitem[Lam \emph{et~al.}(2014)Lam, Cummings, Lin, and Kondic]{Lam2014}
M.~A. Lam, L.~J. Cummings, T.-S. Lin and L.~Kondic, \emph{J Eng. Math.}, 2014,
  \textbf{94}, 97\relax
\mciteBstWouldAddEndPuncttrue
\mciteSetBstMidEndSepPunct{\mcitedefaultmidpunct}
{\mcitedefaultendpunct}{\mcitedefaultseppunct}\relax
\EndOfBibitem
\bibitem[Lam \emph{et~al.}(2015)Lam, Cummings, Lin, and Kondic]{Lam2015}
M.~A. Lam, L.~J. Cummings, T.-S. Lin and L.~Kondic, \emph{Eur. J. Appl. Maths},
  2015, \textbf{25}, 647\relax
\mciteBstWouldAddEndPuncttrue
\mciteSetBstMidEndSepPunct{\mcitedefaultmidpunct}
{\mcitedefaultendpunct}{\mcitedefaultseppunct}\relax
\EndOfBibitem
\bibitem[Mitlin(1993)]{Mitlin1993}
V.~S. Mitlin, \emph{J. Colloid Interface Sci.}, 1993, \textbf{156}, 491\relax
\mciteBstWouldAddEndPuncttrue
\mciteSetBstMidEndSepPunct{\mcitedefaultmidpunct}
{\mcitedefaultendpunct}{\mcitedefaultseppunct}\relax
\EndOfBibitem
\bibitem[Thiele \emph{et~al.}(2016)Thiele, Archer, and Pismen]{ThAP2016prf}
U.~Thiele, A.~Archer and L.~Pismen, \emph{Phys. Rev. Fluids}, 2016, \textbf{1},
  083903\relax
\mciteBstWouldAddEndPuncttrue
\mciteSetBstMidEndSepPunct{\mcitedefaultmidpunct}
{\mcitedefaultendpunct}{\mcitedefaultseppunct}\relax
\EndOfBibitem
\bibitem[Craster and Matar(2009)]{cm_rmp09}
R.~Craster and O.~Matar, \emph{Rev. Mod. Phys.}, 2009, \textbf{81}, 1131\relax
\mciteBstWouldAddEndPuncttrue
\mciteSetBstMidEndSepPunct{\mcitedefaultmidpunct}
{\mcitedefaultendpunct}{\mcitedefaultseppunct}\relax
\EndOfBibitem
\bibitem[Kneppe \emph{et~al.}(1982)Kneppe, Schneider, and
  Sharma]{kneppe_jcp_1982}
H.~Kneppe, F.~Schneider and N.~Sharma, \emph{J. Chem. Phys.}, 1982,
  \textbf{77}, 3203--3208\relax
\mciteBstWouldAddEndPuncttrue
\mciteSetBstMidEndSepPunct{\mcitedefaultmidpunct}
{\mcitedefaultendpunct}{\mcitedefaultseppunct}\relax
\EndOfBibitem
\bibitem[Skarp \emph{et~al.}(1980)Skarp, Lagerwall, and
  Stebler]{skarp_mclc_1980}
K.~Skarp, S.~Lagerwall and B.~Stebler, \emph{Mol. Cryst. Liq. Cryst.}, 1980,
  \textbf{60}, 215--236\relax
\mciteBstWouldAddEndPuncttrue
\mciteSetBstMidEndSepPunct{\mcitedefaultmidpunct}
{\mcitedefaultendpunct}{\mcitedefaultseppunct}\relax
\EndOfBibitem
\end{mcitethebibliography}
\bibliographystyle{rsc_ref.bst} 
\end{document}